\documentclass[lettersize,journal]{IEEEtran}
\usepackage{amsmath,amsfonts}
\usepackage{algorithmic}
\usepackage{algorithm}
\usepackage{array}
\usepackage[caption=false,font=normalsize,labelfont=sf,textfont=sf]{subfig}
\usepackage{textcomp}
\usepackage{stfloats}
\usepackage{url}
\usepackage{verbatim}
\usepackage{graphicx}
\usepackage{cite}
\begin{document}

\title{PowerSkel: A Device-Free Framework Using CSI Signal for Human Skeleton Estimation \\ in Power Station}

\author{Cunyi Yin, Xiren Miao, Jing Chen, 
Hao Jiang,
Jianfei Yang,
Yunjiao Zhou, \\
Min Wu~\IEEEmembership{Senior Member,~IEEE}, and Zhenghua Chen~\IEEEmembership{Senior Member,~IEEE}

\thanks{This work was supported by the Natural Science Foundation of Fujian Province, China, grant number 2022J01566 and the China Scholarship Council, China, grant number 202206650012. (Corresponding author: Jing Chen (chenj@fzu.edu.cn).)

C. Yin, X. Miao, J. Chen and H. Jiang are with the College of Electrical Engineering and Automation, Fuzhou University, Fuzhou, China (cunyiyin1125@gmail.com; mxr@fzu.edu.cn; chenj@fzu.edu.cn; jiangh@fzu.edu.cn), J. Yang and Y. Zhou are with School of Electrical and Electronics Engineering, Nanyang Technological University, Singapore (yunjiao001@e.ntu.edu.sg; yang0478@e.ntu.edu.sg),
M. Wu and Z. Chen are with the Institute for Infocomm Research, Agency for Science, Technology and Research, Singapore (wumin@i2r.a-star.edu.sg, chen0832@e.ntu.edu.sg)


}

}

\markboth{ }%
{Shell \MakeLowercase{\textit{et al.}}: A Sample Article Using IEEEtran.cls for IEEE Journals}


\maketitle

\begin{abstract}
Safety monitoring of power operations in power stations is crucial for preventing accidents and ensuring stable power supply. 
However, conventional methods such as wearable devices and video surveillance have limitations such as high cost, dependence on light, and visual blind spots.
WiFi-based human pose estimation is a suitable method for monitoring power operations due to its low cost, device-free, and robustness to various illumination conditions.
In this paper, a novel Channel State Information (CSI)-based pose estimation framework, namely PowerSkel, is developed to address these challenges. 
PowerSkel utilizes self-developed CSI sensors to form a mutual sensing network and constructs a CSI acquisition scheme specialized for power scenarios. 
It significantly reduces the deployment cost and complexity compared to the existing solutions. 
To reduce interference with CSI in the electricity scenario, a sparse adaptive filtering algorithm is designed to preprocess the CSI. CKDformer, a knowledge distillation network based on collaborative learning and self-attention, is proposed to extract the features from CSI and establish the mapping relationship between CSI and keypoints. 
The experiments are conducted in a real-world power station, and the results show that the PowerSkel achieves high performance with a PCK@50 of 96.27\%, and realizes a significant visualization on pose estimation, even in dark environments. 
Our work provides a novel low-cost and high-precision pose estimation solution for power operation.
\end{abstract}


\section{Introduction}
\IEEEPARstart{T}{he} power system is a crucial infrastructure for economic development worldwide. 
Electric power substations are essential components of the power system, which include indoor transmission, indoor distribution, and switching substations~\cite{bashkari2020outage}. 
Regular inspection and maintenance are vital to ensure a safe and stable power supply. 
The most reliable and effective way to inspect substations is through personnel on-site. 
However, ensuring personnel safety during the inspection and maintenance process is a complex problem that the power industry faces. 
Mistakes and irregularities in the inspection process can result in personal injury and economic losses~\cite{kianoush2016device}.

Extensive research has been done to enhance the safety of individuals who perform power station inspections. 
To achieve this, more advanced monitoring and management strategies have been developed~\cite{lu2017mobile, yang2022vision}. 
Currently, the most common approaches to monitoring personnel safety in power station settings are wearable devices and video surveillance technology~\cite{zhang2022online, yin2020theoretical}. 
However, wearable devices are inconvenient for staff, while lighting conditions and blind spots in camera vision limit video surveillance technology. 
Furthermore, both methods are costly. 
Therefore, there is a need for a low-cost, device-free, and light condition-independent human pose estimation method for indoor power station inspections.

Lately, there has been notable interest in device-free methods for detecting humans~\cite{yin2021device, yin2023human}.
One of the methods that have been receiving attention is WiFi-based sensing technology. 
It is advantageous due to its low cost, ease of deployment, and ability to function without line-of-sight~\cite{chen2021low}. 
Additionally, WiFi can capture fine-grained Channel State Information (CSI), providing better activity features, as demonstrated in studies like~\cite{wang2021fallviewer, khan2020differential}. 
Human pose estimation, which involves inferring a human skeletal model from input signals, has numerous applications, as discussed in~\cite{cao2022joint}. 
Recently, a novel approach using WiFi for human pose estimation has emerged. 
The method involves cross-modal annotation and supervision of CSI during the training phase, enabling the raw CSI to estimate human pose accurately. 
Utilizing CSI for human pose estimation is particularly useful in poorly lit environments. 

Utilizing human pose estimation for power operation monitoring via CSI poses several challenges.
Firstly, current signal acquisition in CSI involves wired connections to the host computer, impractical and costly for power stations with spatial limitations. 
Secondly, power operations are characterized by global and local features, rendering them more complex than typical human activity classification tasks. 
Thirdly, the intricate CSI inherent in power operations poses a significant challenge for accurate pose estimation, particularly given the heightened dimensions and pose features. 
Existing feature extraction methods struggle to effectively handle these large and intricate CSI dimensions.

To address these issues, a novel CSI acquisition method is proposed to meet the needs for power operations, i.e., easy to deploy, low-cost, and unaffected by lighting conditions and line of sight. 
The data acquisition for power operations endows CSI with unique attributes.
The CSI acquisition for power operation establishes a CSI mutual sensing network (mutual matrix) among multiple sensors to extract comprehensive CSI information. 
The mutual sensing matrix incorporates CSI data from all sensors, providing a mutual matrix with global information. 
Each sensor engages in mutual sniffing, forming a sniffing path matrix (path matrix) with every other sensor. 
The path matrix encapsulates local information of the paths it encompasses, constituting the fundamental elements of the mutual matrix.
We leverage the structure of convolution combined with self-attention to balance both global and local features of CSI for more complex human pose estimation. 
Inspired by the success of Knowledge Distillation (KD) in computer vision, KD can dynamically transfer knowledge between models during the training process to alleviate the single-model-constrained feature extraction problem, thus mining the fine-grained information for information exchange and complementation across channels~\cite{gou2021knowledge}.

To this end, a power operation sensing framework utilizing CSI, PowerSkel, is developed based on ESP 32 IoT SoC. 
The CSI sensors enable sniffing each other and uploading data through wireless communication in PowerSkel, which is easy to deploy while acquiring rich CSI. 
The Kinect is utilized in the training stage to obtain accurate visual skeletal coordinates for CSI annotation and training. 
To improve the restriction of a single model, a CSI-based pose estimation model, CKDformer, is proposed based on collaborative KD, which realizes feature extraction and interaction between CSI channels through multiple model cooperation and knowledge fusion.
Conformer~\cite{gulati2020conformer}, a model based on convolution and self-attention mechanism, is leveraged as the backbone of CKDformer to capture fine-grained local and global features of CSI. 
The mapping relationship between CSI and human posture keypoints is established in the training of CKDformer, and the pure CSI human pose estimation for power operation is finally realized. 
Realistic experiments show that the approach proposed in this paper achieves high accuracy in human pose estimation.

We summarize the contributions of this work as follows:
\begin{itemize}
\item{A power operation posture estimation approach for power station is proposed. 
The CSI is employed for human posture estimation in power scenarios, and a data processing method for CSI in power scenarios is leveraged.}
\item{We design PowerSkel, a novel CSI mutual sensing solution leveraging self-developed CSI sensors, which is the first CSI-based human posture estimation solution for power station. Especially in power station with limited light, the PowerSkel remains to work effectively.}
\item{A knowledge distillation network based on collaborative learning and self-attention mechanism, named CKDformer, is proposed to share the global and local information of the CSI channel among multiple Conformers. The keypoints coordinates of Kinect are leveraged for cross-modal supervision to estimate the human pose in power scenario.}
\end{itemize}

\section{RELATED WORK}
\begin{figure}[t]
\centering
\includegraphics[width=2.8in]{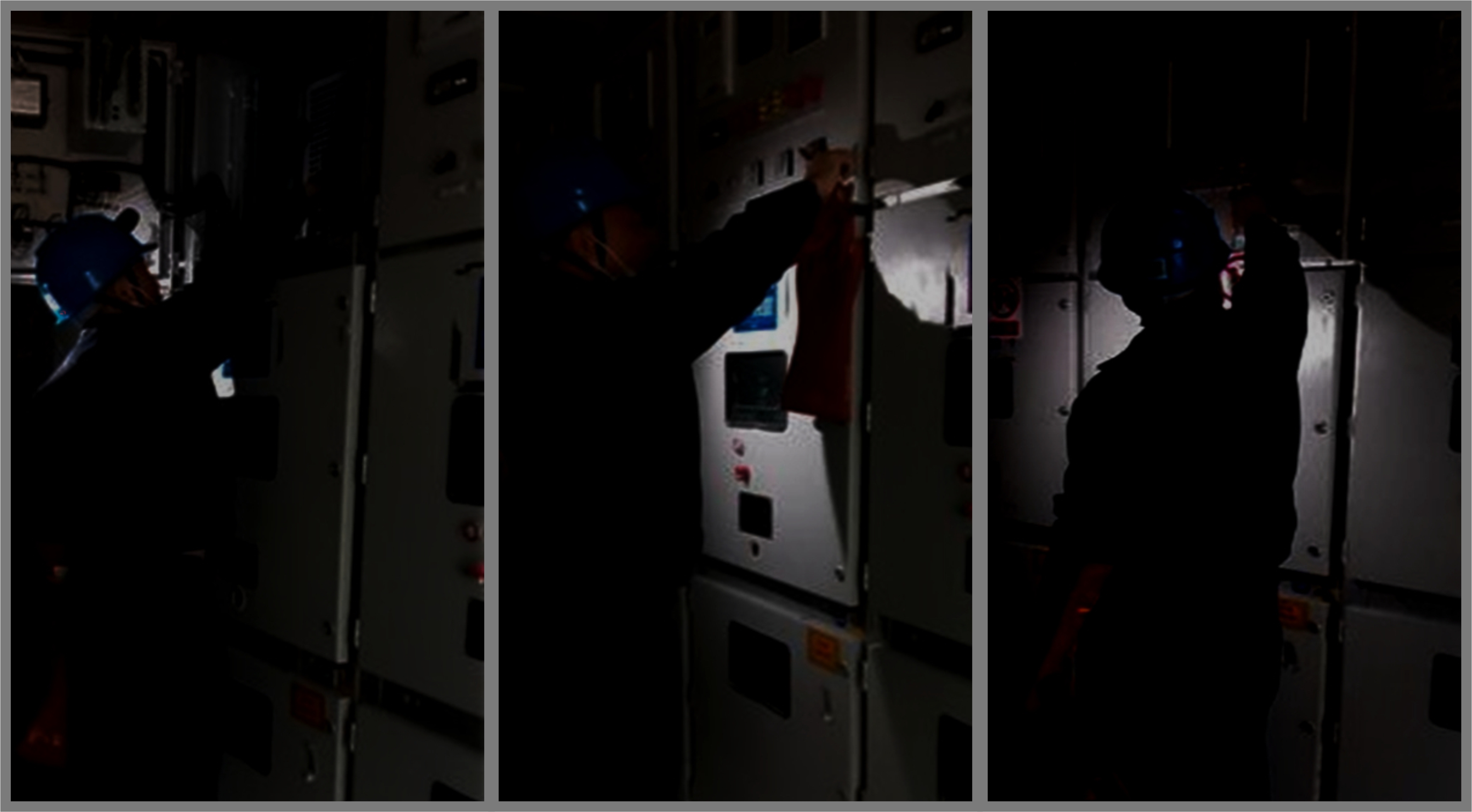}%
\caption{Power operation in dark environments.}
\label{fig_1}
\end{figure}
\subsection{Power Operation Monitoring}
Power operations involve various tasks, from maintenance to operations and restoration work within the power system. 
To ensure the safety of personnel and equipment, it's crucial to follow safety regulations and operating procedures. 
These operations involve electricity and can result in accidents due to poor personnel safety awareness and non-standardized procedures~\cite{yin2020theoretical}. 
Monitoring the safety of power operators is critical. 
Wearable devices and vision-based technology can assist in achieving this goal. 
One way to monitor power operators is through wearable devices such as helmets, clothes, and bracelets that collect information on human pose, movement, and position. 
However, this method can be cumbersome for workers, adding additional weight and raising concerns about battery life~\cite{awolusi2018wearable, ometov2021survey}.

Alternatively, visual-based monitoring techniques are more user-friendly. 
Cameras capture images of power stations, and deep learning models recognize power operations. 
Researchers have used deep learning-based object detection to monitor safety at power grid construction sites~\cite{peng2021cory, bo2021skeleton}. 
However, current research has only addressed outdoor scenarios.
These methods rely on light and may not work effectively in low-light environments, especially indoor power stations. 
Fig. \ref{fig_1} displays an image taken during daily operations at a power station where the operator relies on a portable light as the primary source of illumination.
As a result, it's impractical to use vision-based methods to monitor power operations in dark environments. 
The power station urgently needs a light-insensitive, device-free approach for human pose estimation.

\subsection{WiFi-based Human Sensing}
WiFi-based human sensing technology is a cutting-edge method that uses wireless signals to detect human activities~\cite{yang2018device, yang2019learning, yang2022efficientfi, yang2022autofi}. 
WiFi sensing offers numerous benefits, including privacy protection and cost-effective deployment.
Two categories of WiFi-based human sensing technology are pattern-based and model-based methods. 
Pattern-based methods extract features from CSI and use machine learning techniques for identification, while model-based methods establish wireless propagation models to uncover the physical relationship between CSI and human movement, using models for more accurate identification. 

Deep learning has significantly improved WiFi-based human sensing in recent years, enhancing identification accuracy and robustness. Yang et al. introduced SenseFi, a benchmark for evaluating different deep learning models in WiFi-based human sensing, and an open-source library~\cite{yang2023sensefi}. 
However, despite its many benefits, WiFi-based human sensing still faces challenges, including unstable signal quality and environmental influences~\cite{liu2020human}. 
To overcome these challenges, researchers have explored alternative technologies, such as activity-related feature extraction and enhancement to eliminate noise and retain behavior-related details~\cite{liu2019opportunities}. 

\begin{figure*}[!t]
\centering
\includegraphics[width=6.5in]{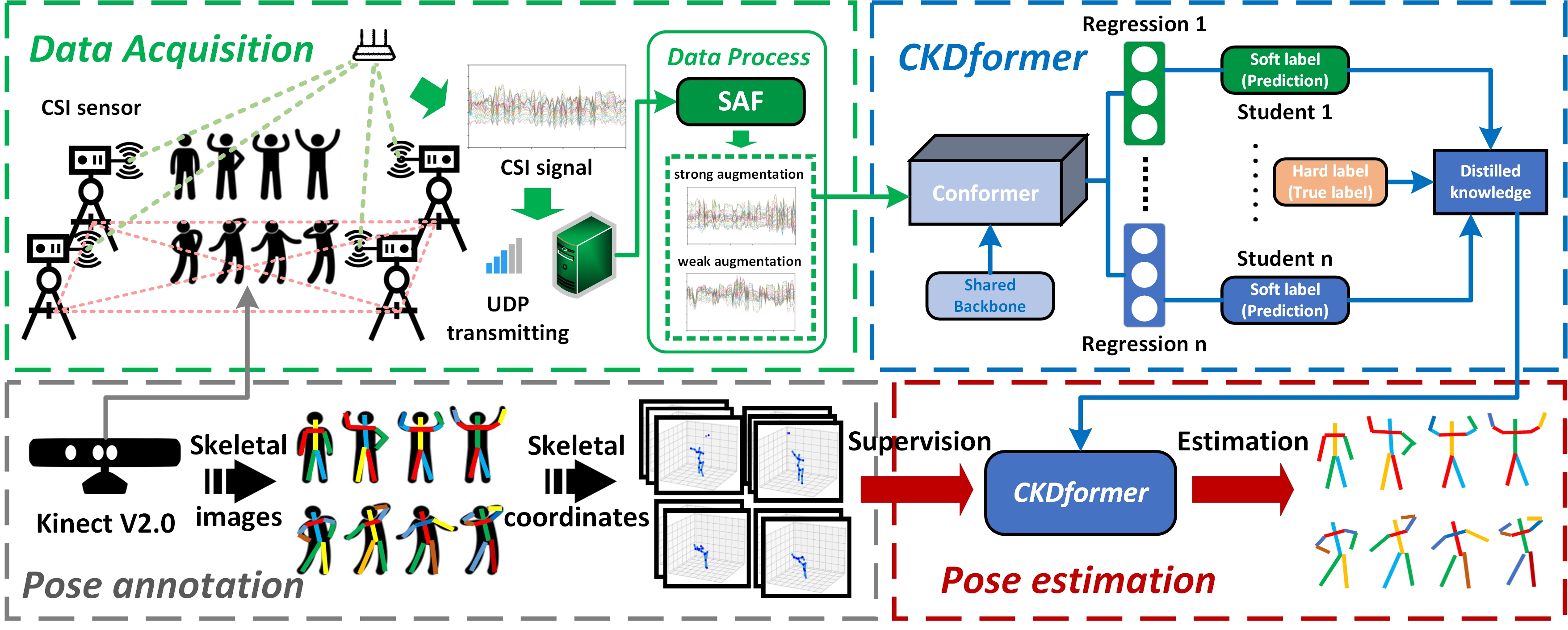}%
\caption{The framework of PowerSkel. PowerSkel establishes wireless communication to transmit CSI and skeletal coordinates to the server. The SAF processes the CSI data for model training. Simultaneously, the keypoints obtained from Kinect serve as a source for cross-modal supervised pose annotation within the CKDformer architecture. This approach enables accurate human pose estimation using pure CSI inputs.}
\label{fig_2}
\end{figure*}

Human pose estimation is another exciting application of WiFi-based human sensing, which infers the skeletal model of a human based on input signals~\cite{wang2021point}. 
While using WiFi for human pose estimation presents challenges due to WiFi signals' poor and unstable quality and the complex mapping relationship between WiFi signals and human pose, researchers have made initial progress in this area~\cite{kim2022human}. 
For instance, Ren et al. utilized the Angle of Arrival spectrum of WiFi signals reflected from the human body to locate different body parts and estimate human pose~\cite{ren20223d}. 
Additionally, Wang and Huang et al. proposed WiSPPN and WiMose, respectively, both utilizing commercial WiFi devices to collect CSI and employ visual models for supervised learning~\cite{wang2019can, wang2019joint}. 
Human pose estimation has given rise to new WiFi applications, such as virtual and augmented reality, that require high accuracy and low latency. 
Yang et al. proposed MetaFi, a device-free pose estimation system for Metaverse virtual character simulation using ordinary WiFi devices~\cite{yang2022metafi}.
MetaFi++ is an improved version of MetaFi, which uses Transformers to extract and fuse human pose information in WiFi signals and improves pose estimation accuracy and robustness~\cite{zhou2023metafi++}.
However, no research on power operations using CSI in power scenarios exists. 
Due to the particularity of the power scenarios and power operations, the previous methods are difficult to reuse here. Therefore, a specialized system and method are needed to accommodate power scenarios.

\subsection{Knowledge Distillation}
Knowledge Distillation (KD) compresses and improves models by having smaller models learn from larger ones~\cite{gou2021knowledge}. 
KD has been successfully applied in computer vision tasks, improving performance and reducing costs~\cite{chen2019knowledge, kang2021instance, zheng2022localization}. 
KD has also been used in vision-based human pose estimation tasks. 
Different teacher networks provide pose structure information in these methods, and techniques~\cite{li2021online, zhao2021orderly}. 
However, these methods must pre-train a complex teacher network with feature aggregation units to enhance detection. Collaborative learning is a promising solution for CSI, involving multiple models with different architectures and capacities that transfer knowledge dynamically~\cite{guo2020online}.


\subsection{Transformer Network}
The Transformer, initially designed for Natural Language Processing, has gained widespread adoption in computer vision tasks such as image classification, target detection, and semantic segmentation~\cite{vaswani2017attention, chen2021crossvit, tian2019fcos, zhao2020exploring}. 
Its self-attention mechanism efficiently captures long-range dependencies, offering scalable modeling capabilities. 
In vision-based human pose estimation, it has demonstrated strong performance. 
For instance, it has been used for recognizing human activities in pose sequences, personalizing pose estimation across variations, modeling hand joint dependencies, and leveraging WiFi-based CSI data for finer-grained feature extraction~\cite{mazzia2022action, li2021test, huang2020hand, zhou2023metafi++}. 
However, prior approaches underutilized the potential of CSI in capturing both global and local feature distributions.
The Conformer's efficient processing of both global and local data stands out~\cite{gulati2020conformer}. 
Its adept balance of these features aligns well with the distribution of CSI signals in power operations, offering insights for extracting CSI features of power operations. The encapsulation of global information is a higher-level abstraction that can be informative to a certain extent but may lack the granularity required to discern nuanced variations among individual sensors. Especially in complex scenarios, excessive generalized information can result in models that struggle to identify comprehensive information in distinct local features. Relying solely on local information proves insufficient in capturing the nuanced characteristics of comprehensive actions. Consequently, striking a balance between global and local perspectives becomes crucial for achieving robustness and accuracy in CSI-based human pose estimation models.
In this work, we aim to use the Conformer mechanism to construct the relationship between global and local features of CSI and to mine the fine-grained information of CSI deeply.

\section{Methodology}
\subsection{PowerSkel System Design}
The PowerSkel framework consists of four key components: data acquisition and processing, pose annotation, model training, and pose estimation, as shown in Fig. \ref{fig_2}. 
In the data acquisition stage, multiple CSI sensors and a Kinect are deployed in the power station, forming a collaborative sensing network. 
These sensors send data wirelessly to the server via UDP, ensuring synchronization between each CSI frame and an image frame containing a pose. 
The data is processed by Sparse Adaptive Filtering (SAF), which is tailored for power scenarios to enhance CSI features by suppressing noise. 
The processed CSI serves as training data, while image-based keypoints serve as labels for model training. 
The Conformer-based model, CKDformer, uses collaborative KD to map CSI and pose data, using the Conformer as the core architecture. 
Channel features and the relationship are shared among different student models. 
A novel Optimal Transport (OT)-based distillation loss facilitates knowledge exchange among student models. 
The framework enables precise human pose estimation from pure CSI inputs only.

\subsection{Data Processing}
Power stations generate electromagnetic fields with frequencies ranging from tens to hundreds of hertz. 
However, it's important to note that the operating frequency of CSI is 2.4 GHz~\cite{perera2017simultaneous}. 
While the power station environment is not expected to significantly impact the CSI, the potential influence of partial discharges and corona generated by other power equipment is still worth considering. 
To address this, our paper introduces the Sparse Adaptive Filtering Algorithm (SAF) to reconstruct the CSI during power operations. 
The core concept of the algorithm is to utilize the sparsity of the input CSI signal within a specific dictionary and update the filter coefficients accordingly. 
By constructing the dictionary from delayed versions of the input signal and obtaining the sparse representation through solving a least squares problem with a sparsity constraint, SAF can effectively update the filter coefficients using a gradient descent method and achieve a low mean squared error.

\begin{algorithm}[t]
    \renewcommand{\algorithmicrequire}{\textbf{Input:}}
    \renewcommand{\algorithmicensure}{\textbf{Output:}}
    \caption{Sparse Adaptive Filtering}
    \label{alg2}
    \begin{algorithmic}[1]  
        \REQUIRE Input CSI signal $x_{in}$, the number of training set CSI samples $N$, step size $\mu$
        \ENSURE Reconstructed CSI $x_r$
        \STATE Initialization: $x_{in}$, filter coefficient $h$;
        \STATE Compute the shared dictionary $A$ using Equation 2; 
        \FOR {$n=0$ to $N$}
        \STATE Compute the sparse representation $s$ using the least square method using Equation 1;
        \STATE Compute the gradient $\nabla$ to update filter coefficient using Equation 3;
        \STATE Update the filter coefficients $h$ using gradient descent as shown in Equation 4 with the step size $\mu$;
        \STATE Reconstruct CSI with dictionaries and sparse representations using Equation 5; 
        \ENDFOR
    \end{algorithmic}
\end{algorithm}
As mentioned in Section I, multiple CSI sensors sniff each other. Each sensor can acquire $f$ subcarrier data from different sensors. $m$ sensors capture $e$ sniffing paths in sniffing each other, and the sensors are cannot sniff themselves, the $e$ can be express as: $e = m \times (m-1)$. 
For the input CSI signal, CSI can be denoted as $x_s\in\mathbb{R}^{N \times e \times f}$, where $N$ represents the number of CSI samples of training set.
The proposed algorithm is described as follows:

For each CSI signal $x_{in}\in\mathbb{R}^{e \times f}$ is flattened into a one-dimensional column vector $d\in\mathbb{R}^{k}$ for matrix multiplication with the dictionary, where $k = e \times f$.
Flattening is traversing all the elements of each row of the matrix in order, adding each element to the one-dimensional vector accordingly.
The sparse representation $s \in\mathbb{R}^{k}$ of $x_{in}$ is a column vector and it computed using least squares as follows:
\begin{equation}
s=\operatorname*{argmin}_{s} ||A s-d||_{2}^{2},
\end{equation}
where $A\in\mathbb{R}^{k \times k}$ is the shared dictionary. A cyclic shift on $d$ to construct the shared dictionary $A$. Specifically, the last element in the column vector $d$ is moved to the first position each time, and then the moved $d$ is utilized as a column in the dictionary. So, each dictionary column is a loop-shifted version of $d$, and it can be expressed as:
\begin{equation}
A = [\operatorname*{roll}(d, 0), \operatorname{roll}(d, 1), \ldots, \operatorname{roll}(d, k-1)].
\end{equation}
where roll($d, j$) represents the shifting on $j$-th element of the column vector $d$, and $k$ is the length of the $d$. It can obtain the $k \times k$ shared dictionary $A$ by performing the preceding operations.
The gradient is calculated to update the filter coefficients, which can be obtained by:
\begin{equation}
\nabla=2A^{\top}(A h-d),
\end{equation}
where $\nabla\in\mathbb{R}^{k}$ is the gradient, $h\in\mathbb{R}^{k}$ is a column vector representing the filter coefficients, which adjust signal amplitude and frequency during filtering, preserving important information. The filter coefficients are updated using the gradient descent method and can be expressed as:
\begin{equation}
h_{n}=h_{o}-\mu\nabla,
\end{equation}
where $h_{n}$ is the new filter coefficients, $h_{o}$ is the old filter coefficients, $\mu$ is the step size. 
The CSI is reconstructed using the dictionary and the sparse representation, the reconstructed CSI $x_{r}\in\mathbb{R}^{k}$ is computed by:
\begin{equation}
x_r = A s.
\end{equation}
The SAF process is summarized in Algorithm 1.
The CSI is handled and duplicated, generating two identical duplicates. 
These duplicates are then subjected to strong and weak augmentation before being introduced into CKDfomer. 
Strong augmentation involves injecting random noise into the CSI, heightening interference and noise levels. 
On the other hand, weak augmentation includes cyclic panning of the CSI, enhancing its diversity. 
These procedures result in improved overall performance and robustness for PowerSkel.

\begin{figure*}[!t]
\centering
\includegraphics[width=7.0in]{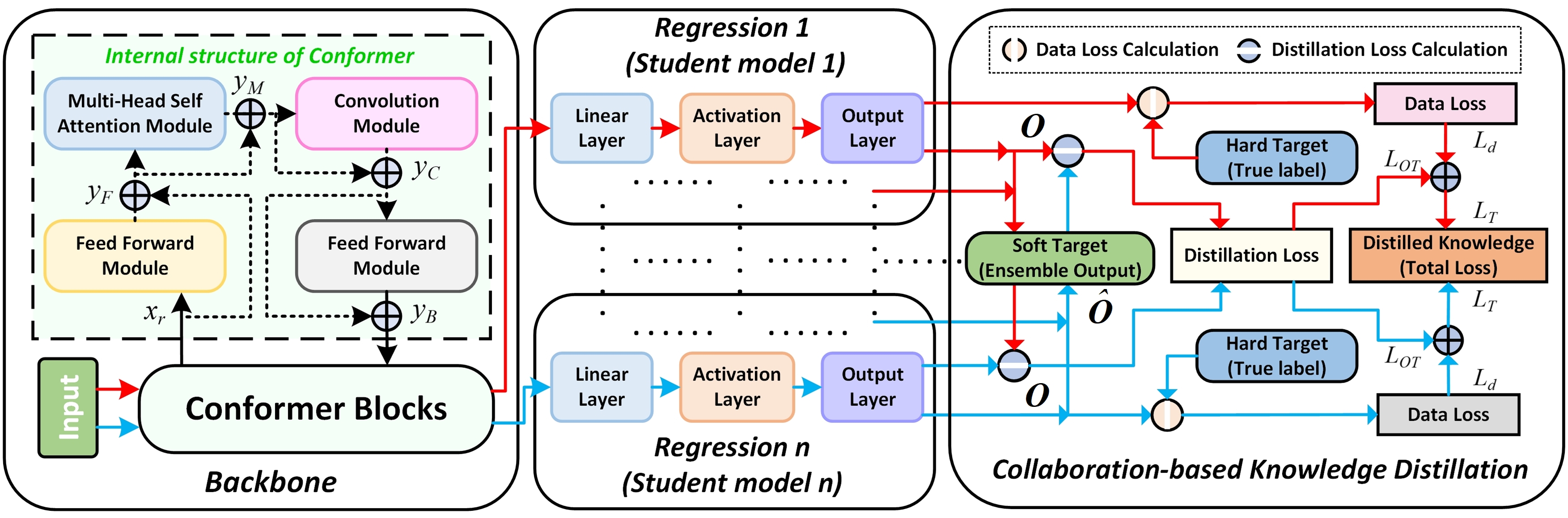}%
\caption{The structure of CKDformer. The CKDformer architecture employs Conformer as a shared backbone to extract local and global CSI features. These extracted features are utilized to calculate distillation loss through collaborative learning. The combination of distillation loss with data loss enhances the regression results of the student models, minimizing the gap between student models and ultimately achieving accurate power operation pose estimation.}
\label{fig_4}
\end{figure*}
\subsection{CKDformer: Collaborative Knowledge Distillation Conformer Network}
PowerSkel is a novel framework for training CSI to comprehend human poses visually and utilize this knowledge for direct pose estimation. 
The labels for CSI are obtained from human pose keypoints captured by the Kinect V2.0. 
To achieve this, a CKDformer model is employed, which maps the CSI to the coordinates of the keypoints. 
Different from previous approaches, CKDformer adopts collaborative learning by dynamically generating a virtual teacher model that weighs predictions from multiple student models. 
The collaborative learning strategy enhances the quality and diversity of soft targets, facilitating better knowledge sharing. 
The architecture of CKDformer, as illustrated in Fig. \ref{fig_4}, incorporates the Conformer as a shared backbone, which extracts both local and global CSI features. 
These features then undergo separate regressions to estimate the pose coordinates. 
The main objective of CKDformer is to achieve accurate pose estimation based on CSI by continuously minimizing the error between the predicted CSI keypoints and the ground truth.

CKDformer is a network composed of multiple student models, where each student model receives individually augmented CSI. 
Conformer~\cite{gulati2020conformer} demonstrates exceptional processing capabilities for global and local data information, providing a distinct advantage in extracting features from CSI signals in power operations. 
This feature extraction process is crucial for accurately estimation and analyzing complex power operations.
The Conformer serves as the foundation of the student model and effectively extracts spatial and local features from the CSI. 
It consists of an input and output layer, along with several Conformer blocks, each of which contains four sub-modules: the Feedforward Module ($\mathrm{FF}$), the Multi-head Self-Attention Module ($\mathrm{MSA}$), the Convolution Module ($\mathrm{Conv}$). 
The input to the Conformer block is the SAF-processed reconstructed CSI, $x_{r}\in\mathbb{R}^{k}$.
The outputs of Feedforward Module ($\mathrm{FF}$) are represented by the $y_F$ as:
\begin{equation}
y_F=x_r + {\frac{1}{2}}\mathrm{F F}(x_r),
\end{equation}
where $y_{F}\in\mathbb{R}^{k}$, the $\mathrm{FF}$ increases the nonlinear capability of the network and transforms different feature dimensions. The FF is computed by:
\begin{equation}
\mathrm{FF}(x_r)=\sigma(W_{F} x_r+b_F),
\end{equation}
where, $W_{F}$ is the weight matrix, $b_F$ is the bias vector, and $\sigma$ is the activation function. 
The outputs of Multi-head Self-Attention Module ($\mathrm{MSA}$) can be represented as follows:
\begin{equation}
y_M=y_F+\mathrm{MSA}(y_F),
\end{equation}
where $y_{M}\in\mathbb{R}^{k}$, the $\mathrm{MSA}$ receives the $y_F$ as input. The $\mathrm{MSA}$ is calculated by Eqs. (9) and (10), as follows:
\begin{equation}
\mathrm{M S A}(Q, K, V)=\lbrack head_{1},...,head_{n}]{\cal W}_{O},
\end{equation}
where ${\cal W}_{O}\in\mathbb{R}^{k \times k}$ is the learnable parameter, $h e a d_{i}\ =\ \mathrm{Attention}(y_F{\mathcal W}_{Q}^{i},y_F{\mathcal W}_{K}^{i},y_F{\mathcal W}_{V}^{i})$, and the attention mechanism is denoted as:
\begin{align}
    \mathrm{Attention}(Q,K,V) &= \mathrm{Attention}(y_F{\mathcal W}_{Q},y_F{\mathcal W}_{K},y_F{\mathcal W}_{V}) \nonumber \\
    &= \mathrm{Softmax}\left(\frac{Q K^{\top}}{\sqrt{d_{k}}}\right)V,
\end{align}
where $d_k=k/n$ is the dimension of the key vector, $Q$, $K$, $V$ are the query, key and value matrices for the head, ${\mathcal W}_{Q}\in\mathbb{R}^{k \times k/n}$, ${\mathcal W}_{K}\in\mathbb{R}^{k \times k/n}$, ${\mathcal W}_{V}\in\mathbb{R}^{k \times k/n}$ are projection parameters, $n$ is the number of $h e a d_{i}$.
The $\mathrm{MSA}$ adeptly captures global connections, while $\mathrm{Conv}$ reveals detailed features by detecting local correlations.
The output of Convolution Module ($\mathrm{Conv}$) is expressed mathematically:
\begin{equation}
y_C=y_M+\mathrm{C o n v}(y_M),
\end{equation}
where the Conv computes input $y_M$ as follows:
\begin{equation}
\mathrm{Conv}(y_M)=ReLU\left(\sum_{i=1}^{Ke}W_{C}y_{M}+b_C\right),
\end{equation}
where $Ke$ is the size of the convolution kernel, $W_C$ is the weight of the convolution kernel, $b_C$ is the bias term, and $ReLU$ is activation function.
The output of $\mathrm{FF}$ is obtained by aggregating the $y_C$ and the $\mathrm{FF}$ result of $y_C$, which can be obtained by:
\begin{equation}
y_B=\mathrm{Layernorm}(y_C+{\frac{1}{2}}\mathrm{F F}(y_C)),
\end{equation}
where $y_{B}\in\mathbb{R}^{k}$, $\mathrm{Layernorm}$ is a layer normalization operation. The $\mathrm{FF}$ enhances the nonlinear capability of the network and transforms it into different feature dimensions. 

The student models in CKDformer share a single Conformer backbone, and different regressions are used to predict skeletal coordinates. 
CKDformer eliminates the need to re-train separate backbones for each student model, which significantly reduces training costs compared to per-student model re-training. 
The critical factor in facilitating knowledge sharing in CKDformer is the design of the distillation loss. 
In the KDCL framework, KL divergence measures the disparity between student model outputs and soft targets. 
However, KL divergence may not be the best approach for regression tasks that involve numerical results. 
Additionally, KL divergence may not adequately account for the geometric structure and deformation in the output space, which is crucial for regressions. 
Thus, we use Optimal Transport (OT) as the distillation loss for regressions~\cite{courty2017joint}. 
The OT effectively compares probability distributions by minimizing transportation costs, guided by a chosen distance function. 
By employing OT as the distillation loss, the OT captures nuanced differences and variations between student model outputs and soft targets, resulting in improved alignment in the output space. 
The proposed method enhances the knowledge transfer and alignment between student models and their soft targets in regression tasks.
Formally, let $P$ be the set of probability distributions on $R$, and let $c$:$R\times R\to R_+$ be a non-negative cost function. The optimal transport distance between two distributions $p,q\in P$ is defined as:
\begin{equation}
    W_{c}(p,q)=\inf_{\pi\in\Pi(p,q)}\int_{R\times R}c(z,v)d\pi(z,v),
\end{equation}
where, $\Pi(p,q)$ is the set of joint distributions on $R \times R$ with marginals $p$ and $q$. Intuitively, $\pi(z,v)$ represents the mass transported from $z$ to $v$, and $W_c(p,q)$ represents the minimal total cost of transporting $p$ to $q$. A common choice of the cost function is
$ c(z,v){=}|z-v|^{p} $ for some $p\geq1$, which leads to the $Wasserstein-p$ distance.

In our setting, each student model obtains a vector output $O\in\mathbb{R}^{t}$, where $t$ is the keypoints coordinate of the human body. Calculating the soft target in the proposed online knowledge distillation mechanism can provide guidelines and references for learning student models. 
\begin{algorithm}[h]
    \renewcommand{\algorithmicrequire}{\textbf{Input:}}
    \renewcommand{\algorithmicensure}{\textbf{Output:}}
    \caption{Sinkhorn Algorithm for Distillation Loss}
    \label{alg1}
    \begin{algorithmic}[1]  
        \REQUIRE output of student model $O$, soft target $\hat{O}$, regularization parameter $\epsilon$, the number of elements of the input data $t$, maximum iterations $niter$
        \ENSURE Sinkhorn Loss $L_{OT}$
        \STATE Initialization: Compute cost matrix $C$ using Equation 15;
        \STATE Initialization: Initialize the marginal distributions $\phi$ and $\psi$ with equal weights;
        \STATE Initialization: Set parameters of the Sinkhorn algorithm $thresh$, stopping criterion;
        \STATE Initialization: Set vectors $u$ and $v$ as zero vectors, set error $err$ to 0; 
        \FOR {$iter=0$ to $niter$}
        \STATE Compute updated $u$ and $v$ using Equation 16;
        \STATE Compute error $err$ using Equation 18;
        \STATE If $err$ \textless $thresh$, then break the loop;
        \ENDFOR
        \STATE Compute transport plan $z$ using Equation 19; 
        \STATE Compute Sinkhorn loss $L_{OT}$ using Equation 20;
        \STATE Return $L_{OT}$.
    \end{algorithmic}
\end{algorithm}
By comparing the output $O$ of the student model with the soft target $\hat{O}\in\mathbb{R}^{t}$ and minimizing the loss between them, the prediction of the student model can be made as close as possible to the soft target. 
The student model learns valuable information from the soft target, thus improving its prediction performance. The average of all student model outputs is the soft target $\hat{O}$. The Sinkhorn algorithm calculates the distillation loss $L_{OT}$ between the student model output $O$ and the soft target $\hat{O}$. The student model output $O$ is compared to the soft target $\hat{O}$, and the loss $L_{OT}$ is minimized to minimize the loss between them.
The specific procedure for calculating distillation loss $L_{OT}$ with the Sinkhorn algorithm is listed.
The cost matrix $C$ is first constructed, and each element $C_{i j}$ of the cost matrix $C$ is obtained by computing the Euclidean distance, which can be expressed as:
\begin{equation}
C_{i j}=\left|O_i-\hat{O}_j\right|^{2},
\end{equation}
where $O_i$ and $\hat{O}_j$ are the $i-$th element in the student model output and the $j-$th element in the soft target, respectively.

The vetors $u$ and $v$ are initialized to zero vectors and start iterating. In each iteration, the $u$ and $v$ are updated according to the following formulas:
\begin{equation}
\begin{aligned}
u=\epsilon\left(\log(\phi)-lse(M(u,v))\right)+u,    \\
v=\epsilon\left(\mathrm{log}(\psi)-lse(M(u,v))^{\top}\right)+v, 
\end{aligned}
\end{equation}
where $\phi$ and $\psi$ are two marginal distributions, both are uniform distributions with weights $1/t$, $\epsilon$ is the regularization parameter that controls the accuracy and stability of the Sinkhorn algorithm, $lse$ represents the log-sum-exp operation, $M(u, v)$ is the Normalized Cost Matrix, calculated as:
\begin{equation}
M_{i j}=\frac{-C_{i j}+u_{i}+v_{j}}{\epsilon}.
\end{equation}
The error is calculated as the absolute sum of the differences between the new and old $u$:
\begin{equation}
err=\sum_{i}^{t}|u_{i}-u_{i-1}|,
\end{equation}
where $u_i$ is the $u$ of the current iteration and $u_{i-1}$ is the $u$ of the previous iteration. If $err$ is less than a pre-set threshold $thresh$, the $u$ is considered to converge and stop iterating.
The transport plan $z$ is computed as follows:
\begin{equation}
z=\exp(M(u,v)).
\end{equation}
The Sinkhorn loss i.e., $L_{OT}$ is calculated by:
\begin{equation}
L_{OT}=\sum_{i=1}^t \sum_{j=1}^t z_{i j} C_{i j},
\end{equation}
where $L_{OT}$ measures the distribution difference between the output $O$ of the student model and the soft target $\hat{O}$.
The Sinkhorn algorithm is summarized in Algorithm 2.

Simultaneously, the mean absolute error is employed to compute the data loss for each student model, denoted as $L_d$. 
The $L_d$ signifies the difference between the predicted value of each student model and the actual label of the data. 
Consequently, $L_T$ signifies the total loss, constituting the weighted sum of the data loss and the distillation loss. 
It can be expressed as:
\begin{equation}
L_T=\beta L_d + (1-\beta) L_{OT} ,
\end{equation}
where $\beta$ is a weighting parameter to control the relative importance of data loss and distillation loss in the total loss.

\section{EXPERIMENTS}
\subsection{Setup}
We collected the data in a training room designed for 10 kV distribution purposes, following established electrical work procedures and using appropriate safety gear. 
The data collection involved various critical activities for power system maintenance, such as trolley swinging, cabinet door operation, switch disconnections, upper cabinet door closing, grounding knife gate closing, and hanging signage. 
\begin{figure}[t]
\centering
\includegraphics[width=2.85in]{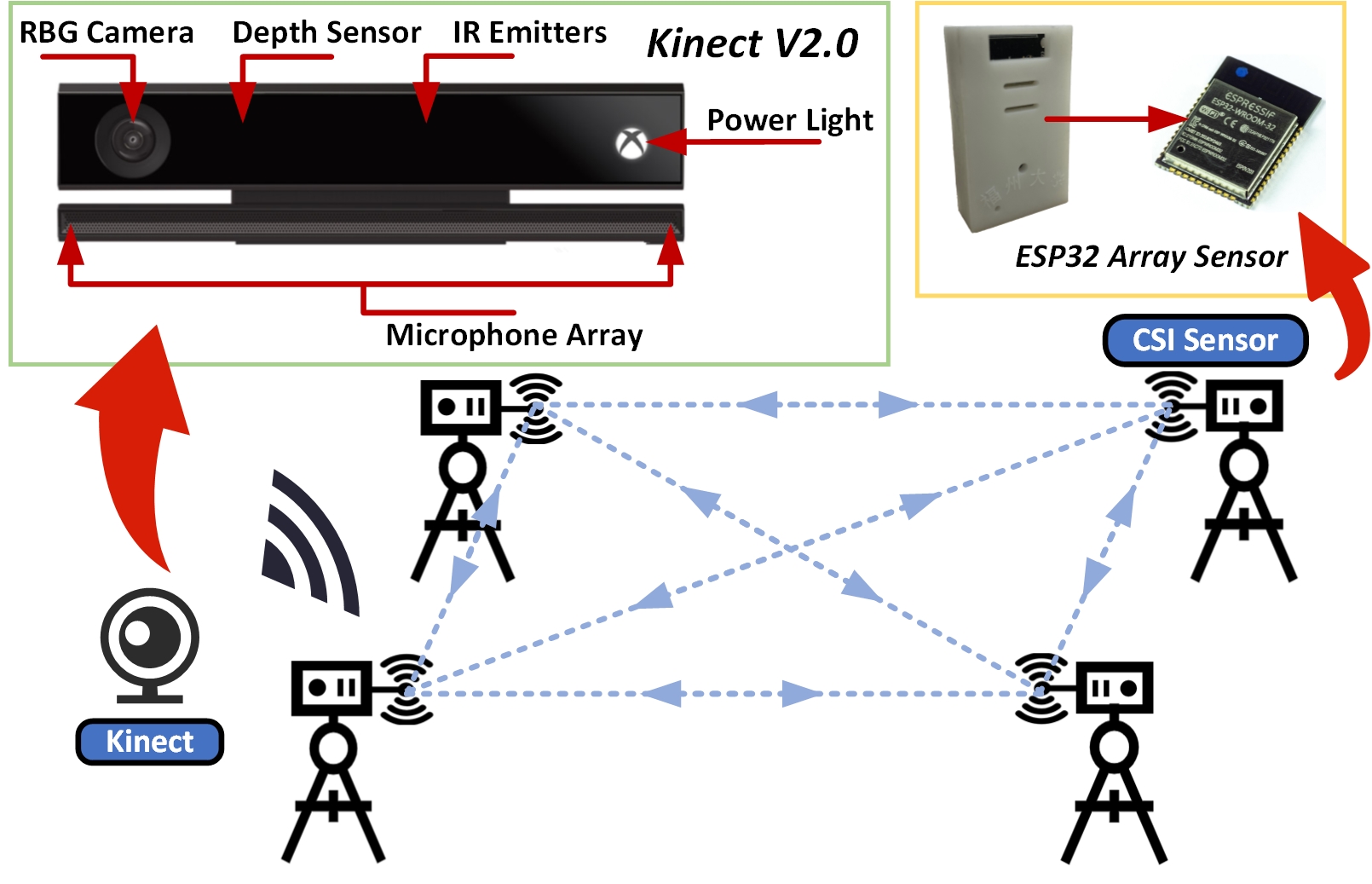}
\caption{Schematic of devices deployment for PowerSkel.}
\label{fig_3}
\end{figure}
The ESP 32 IoT SoC was used to develop the CSI collection device in this paper.
The ESP 32 SoC integrated a 2.4 GHz Wi-Fi and Bluetooth dual-mode MCU, offering low power consumption and stable performance. 
Utilizing the Espressif IoT Development Framework, firmware was developed for the ESP 32, enabling CSI collection. 
The CSI sensor captured CSI from multiple devices and transmitted it to the server via UDP protocol. 
\begin{figure}[!t]
\centering
\includegraphics[width=3.4in]{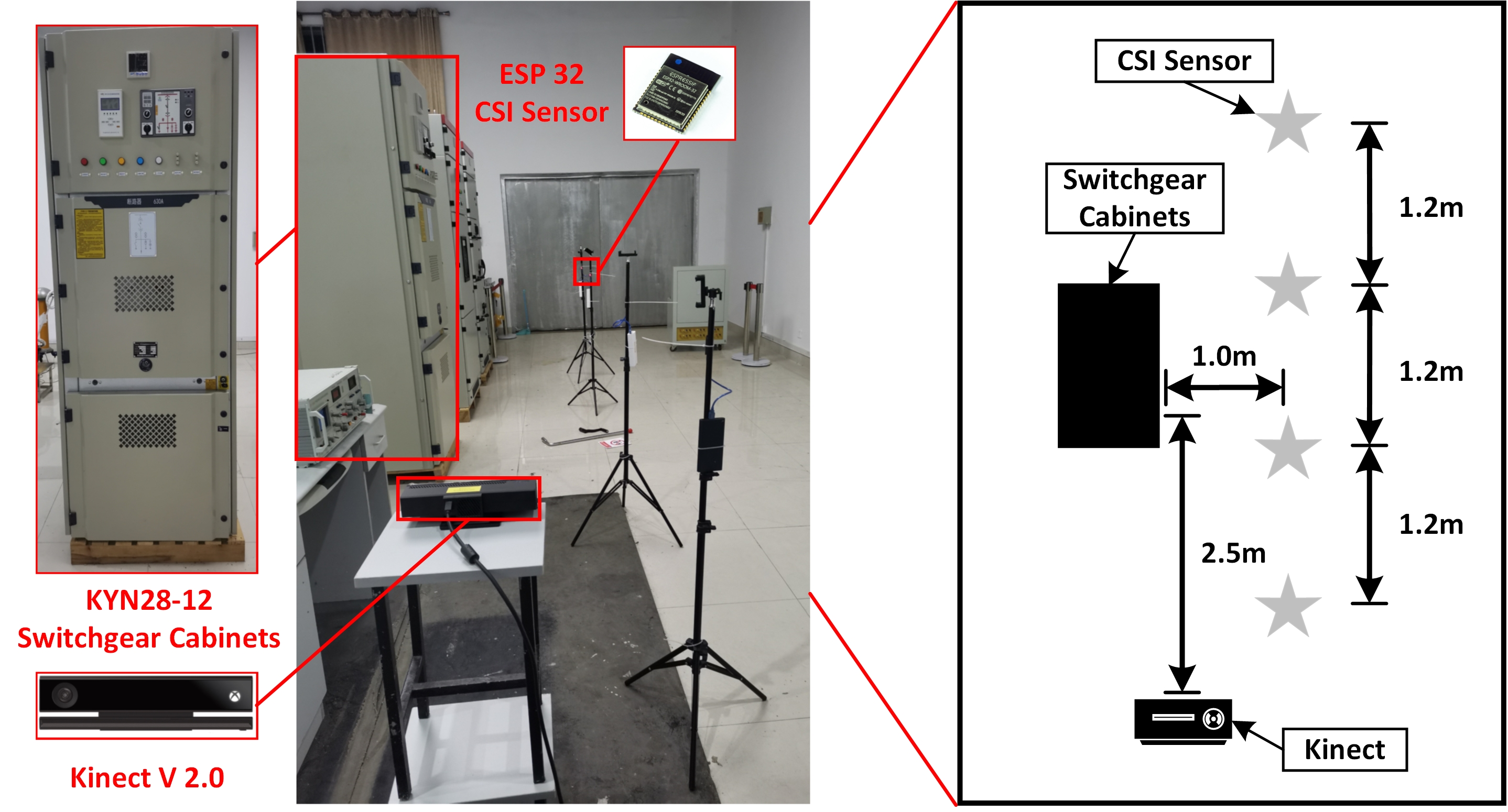}%
\caption{The layout of the experimental environment and Planar graph. These CSI sensors engage in mutual sniffing, serving as both transmitters and receivers.}
\label{fig_5}
\end{figure}
A host program based on the MQTT protocol facilitated remote configuration of CSI sensors. 
The host program enables one-to-multiple CSI acquisition by recognizing Mac addresses from different devices. 
The device was compact, cost-effective, stable, and easily scalable, which was particularly beneficial for power scenarios. 
The Kinect device had an IR Emitter that could perform pose recognition even in low-light environments. 
It also had edge arithmetic power for fast and accurate keypoint extraction. 
The setup provided a benchmark for CSI pose estimation in poorly lit power stations.
Fig. \ref{fig_3} shows the PowerSkel setup, which uses several CSI sensors within a power station. 
Each sensor worked as a sniffer and client at the same time. The sensors formed bidirectional links, creating a mutual sensing network. 
Four CSI sensors and a Kinect V2.0 captured CSI at a sampling rate of 30 Hz. 
The CSI was synchronized with the Kinect data, ensuring corresponding data for each frame of CSI. 

\begin{figure}[!t]
\centering
\includegraphics[width=1.3in]{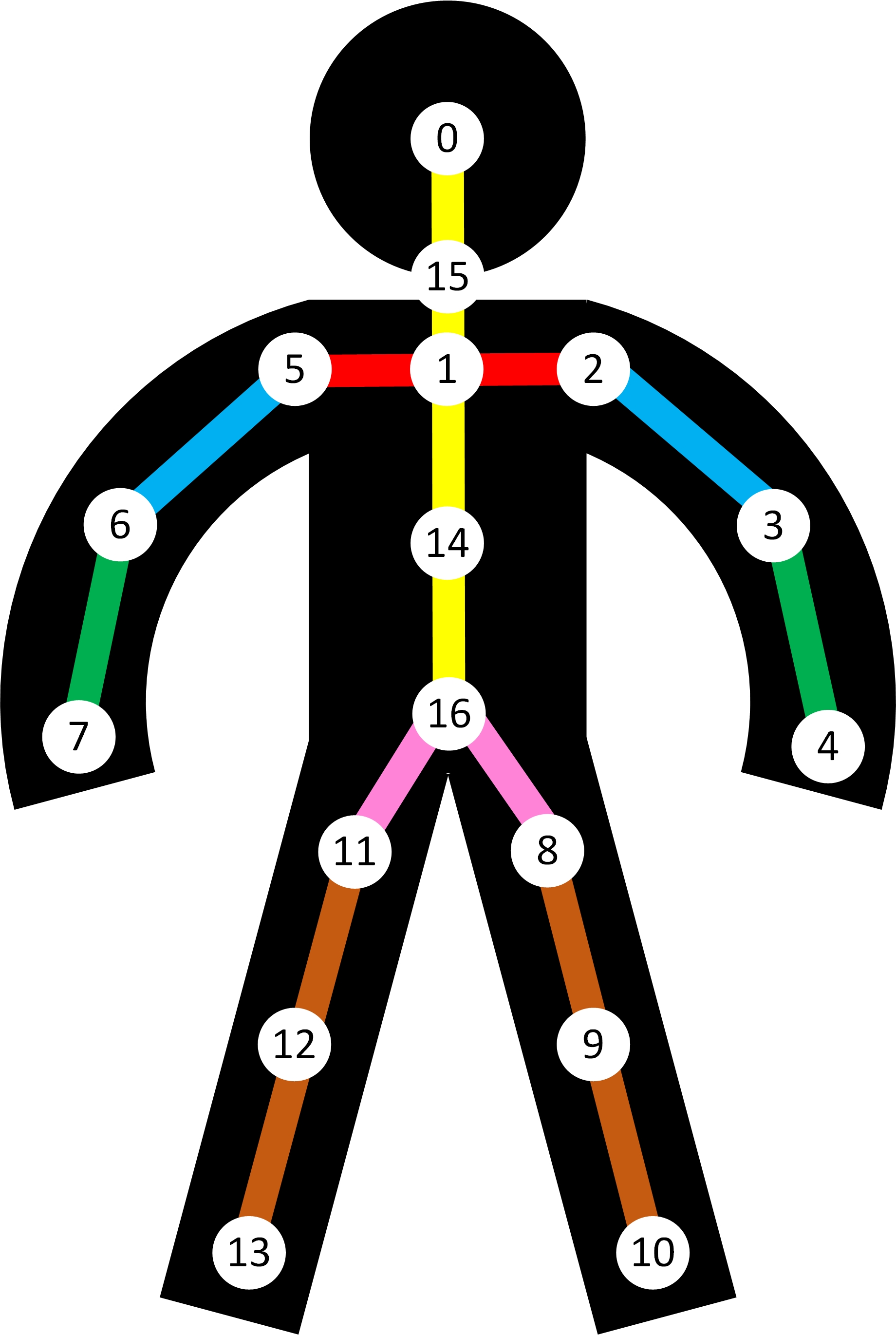}
\caption{Schematic of the 17 keypoints in the experiment~\cite{tolgyessy2021evaluation}.}
\label{fig_6}
\end{figure}

Fig. \ref{fig_5} shows the experimental environment, which consists of multiple KYN28-12 switchgear cabinets. 
The Kinect device was placed 2.5 m from the switchgear and 1.1 m high. 
The sensors were arranged in a straight line, with a spacing of 1.2 m between them, and parallel to the 1.2 m-length switchgear. 
They were located 1 m away from the switchgear and 1.3 m high to ensure safety during operation. 
The sensors sent the CSI matrix to the server via UDP. 
The server processed the CSI, while a Kinect device captured skeletal keypoints as labels. 
The pykinect 2 third-party module in Python was utilized to detect 17 skeletal keypoints. 
These keypoints served as labels for CSI, guiding model training.
Fig. \ref{fig_6} illustrates the 17 keypoints of the human joint skeleton, which provide valuable insights into the movements associated with critical power operation activities~\cite{tolgyessy2021evaluation}.
To evaluate the accuracy of the pose estimation, we use the Percentage of Correct Keypoint (PCK) metric, which is normalized to account for prediction errors~\cite{wang2019person, zhou2023metafi++}, it is defined by:
\begin{equation}
P C K_i @ \alpha=\frac{1}{N} \sum_{i=1}^N I\left(\frac{\left\|p r_i-g r_i\right\|_2^2}{\sqrt[2]{r s^2+l h^2}} \leq \alpha \right)
\end{equation}
where $I$ is the logical indicator, true outputs 1, and false outputs 0. 
$N$ is the number of frames in the test set CSI, $i$ ranges from 0 to 16, and corresponds to each keypoint. 
\begin{table}[t]
\centering
\caption{Results of CKDformer, while L. denotes left \\ and R. denotes right. }
\label{tab1}
\scalebox{0.85}{
\begin{tabular}{c|c c c c c}
\hline
Keypoint          & PCK@10   & PCK@20            & PCK@30   & PCK@40            & PCK@50              \\ \hline
\textbf{Head}              & \textbf{79.62}                        & \textbf{89.45}                        & \textbf{96.84}                        & \textbf{97.97}                        & \textbf{99.12}                        \\
Chest             & 64.27                        & 81.68                        & 91.47                        & 93.53                        & 96.27                        \\
R.Shoulder        & 51.15                        & 81.23                        & 90.38                        & 92.42                        & 94.75                        \\
R.Elbow           & 39.03                        & 72.76                        & 86.64                        & 91.85                        & 92.39                        \\
R.Wrist           & 58.13                        & 85.59                        & 91.22                        & 95.68                        & 96.56                        \\
L.Shoulder        & 72.62                        & 86.81                        & 93.87                        & 96.34                        & 97.93                        \\
L.Elbow           & 75.93                        & 90.8                         & 96.92                        & 98.14                        & 99.54                        \\
L.Wrist           & 67.46                        & 82.68                        & 93.26                        & 96.86                        & 97.96                        \\
R.Hip             & 46.26                        & 73.05                        & 85.45                        & 89.02                        & 91.86                        \\
R.Knee            & 57.25                        & 85.08                        & 89.31                        & 92.47                        & 94.05                        \\
R.Ankle           & 48.23                        & 63.78                        & 76.85                        & 87.79                        & 89.46                        \\
L.Hip             & 72.62                        & 87.41                        & 93.72                        & 97.46                        & 98.58                        \\
L.Knee            & 75.71                        & 90.63                        & 95.53                        & 98.33                        & 98.96                        \\
L.Ankle           & 59.85                        & 85.65                        & 92.74                        & 94.5                         & 95.61                        \\
Abdomen           & 68.68                        & 86.73                        & 93.26                        & 95.71                        & 97.68                        \\
Neck              & 70.15                        & 87.29                        & 94.49                        & 96.92                        & 98.43                        \\
Sacrum            & 72.02                        & 88.65                        & 93.18                        & 95.65                        & 97.52                        \\ \hline
Average           & 63.47       & 83.49          & 91.48       & 94.74          & 96.27        \\ \hline
\end{tabular}
}
\end{table}
$||pr_{i}-gr_{i}||_{2}^{2}$ denotes the Euclidean distance between the estimate and the ground truth, where $pr$ is the predicted results of the predicted skeletal coordinates, and $gr$ is the ground truth, while $\sqrt[2]{r s^2+l h^2}$ denotes the length between right shoulder ($rs$) and the left hip ($lh$), also termed as the length of torso, which is used to normalize the prediction error.
The $\alpha$ is crucial for determining the proximity criterion between the predicted and actual keypoints.

The dataset includes 3 males and 3 females, with distinct training and test data. 
Each sample has $12 \times 51$ data from the CSI sensor as input i.e., it has 12 sniffing paths and each sniffing path has 51 subcarrier signals corresponding to 17 skeletal coordinates (X and Y coordinates) totaling 34 values. In SAF, the parameters $e$ and $f$ are 12 and 51, respectively, and the initial filter coefficient is the vector $h\in\mathbb{R}^{612}$ with element 0, and step size $\mu=500$.
In Sinkhorn algorithm, the parameters involved: regularization parameter $\epsilon=0.01$, the number of elements of the input data $t=34$, maximum number of iterations $niter=100$.
Training data is 25,496, and test data is 5,209. The CKDformer is pytorch-based, with two student models sharing a Conformer backbone. 
Backbone parameters are 612 input dimensions, 4 encoder layers, and 6 multi-heads. 
FF hidden layer dimension is 128, and Conv kernel size is 15.
\begin{figure}[!t]
\centering
\includegraphics[width=3.4in]{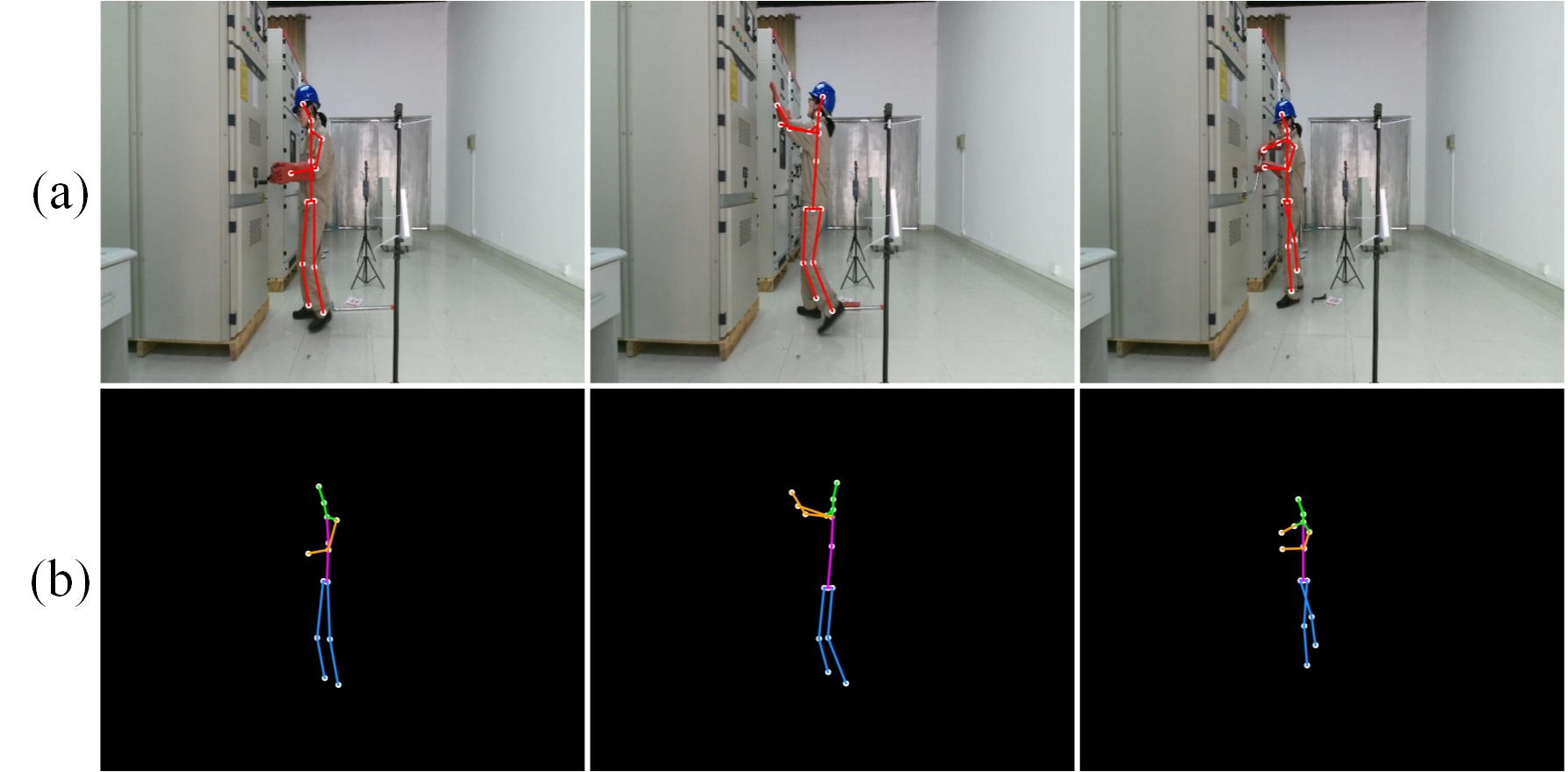}
\caption{Visualization of the human pose landmarks generated by vision model (up) and CSI model (down). (a) Recognition results by Kinect (Ground truth). (b) Estimation of CSI.}
\label{fig_7}
\end{figure}
\begin{figure*}[!t]
\centering
\includegraphics[width=6.8in]{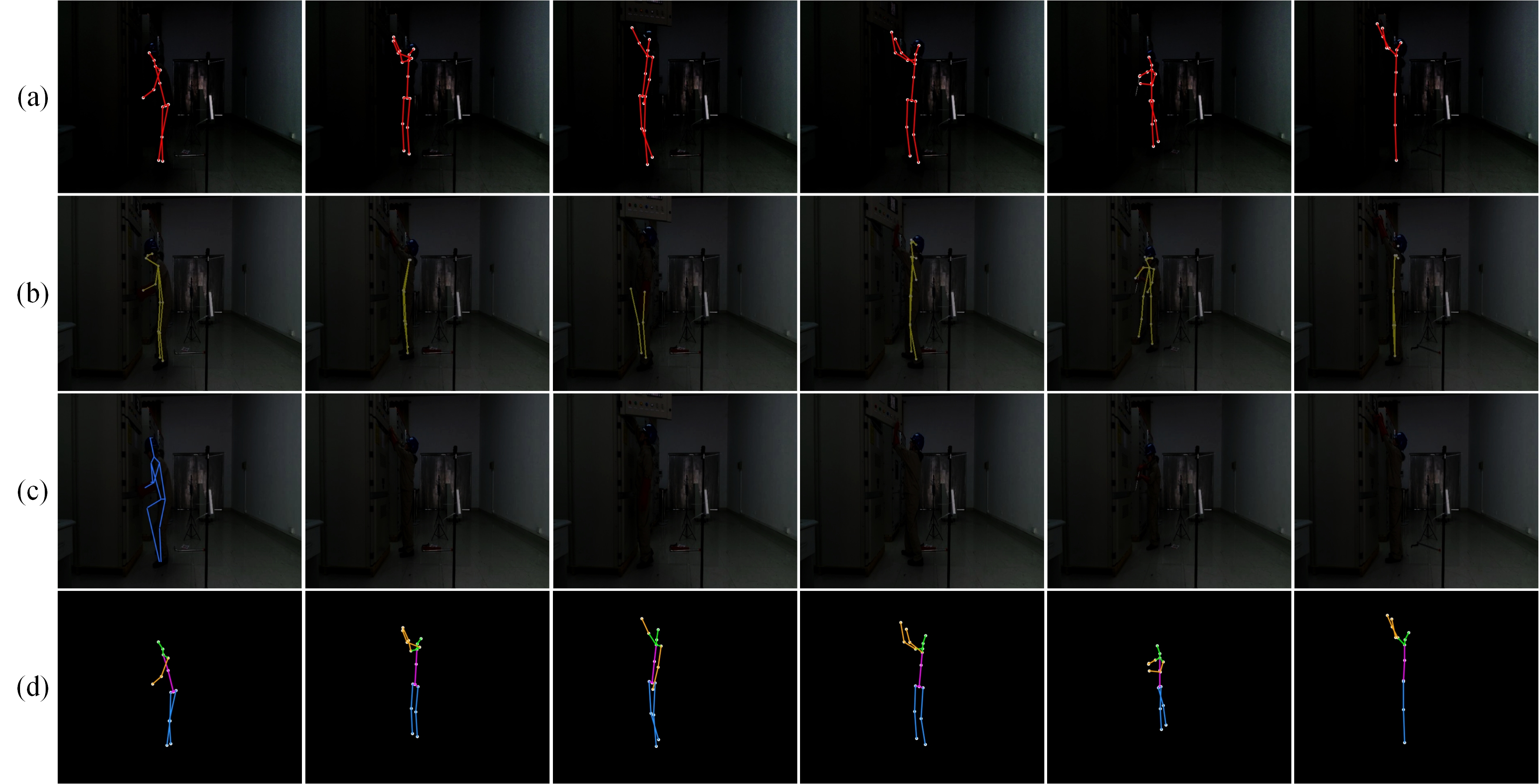}%
\caption{Visual comparison of the estimation with different methods. (a) Ground truth. (b) Estimation of OpenPose. (c) Estimation of MEGVII. \\ (d) Estimation of PowerSkel.}
\label{fig_8}
\end{figure*}
\begin{table}[!h]
\centering
\caption{Cost and functionality comparison of different equipment.}
\label{tab2}
\scalebox{0.85}{
\begin{tabular}{c|cccc}
\hline
Equipment                                                          & \multicolumn{1}{c}{Price(\$)} & \multicolumn{1}{c}{Manufacturer} & \multicolumn{1}{c}{\begin{tabular}[c]{@{}c@{}}Dark or poorly \\ light detection\end{tabular}} & Blind sector \\ \hline
\begin{tabular}[c]{@{}c@{}}Natural light camera\end{tabular} & 30                             & Hikvision                         & -                                                   & $\surd$            \\
\begin{tabular}[c]{@{}c@{}}Night vision camera\end{tabular}  & 70                             & Hikvision                         & $\surd$                                                   & $\surd$            \\
Event camera                                                    & 7800                           & Inivation                         & $\surd$                                                   & $\surd$            \\
Kinect v2                                                       & 300                            & Microsoft                         & $\surd$                                         & $\surd$   \\
Intel 5300NIC                                                   & 400                            & Intel                             & $\surd$                                                   & -            \\
Atheros                                                         & 450                            & TP-Link                           & $\surd$                                                   & -            \\
\textbf{ESP 32 SoC}                                                      & \textbf{3}                              & \textbf{Espressif}                         & \textbf{$\surd$}                                                   & \textbf{-}            \\ \hline
\end{tabular}
}
\end{table}
Two identical regressions follow the backbone, with 2 linear layers and ReLU activation.
A linear output layer has 34 dimensions (17 x and y-coordinates), i.e. the dimension for the output $O$ of the student model and the soft target $\hat{O}$ are $t=34$. 
Training spans 300 epochs, utilizing SGD optimizer, batch size 256, and initial learning rate 0.001. The source code of the proposed CKDformer is available on the open-source GitHub\footnote{https://github.com/power-operation/CKDformer}.

\subsection{Results}
Table I displays the results for the 17 keypoints, showcasing the overall performance of pose estimation. 
The CKDformer demonstrates impressive accuracy in specific and average keypoints pose estimation, achieving a PCK@50 score exceeding 91\% for 16 of 17 keypoints. 
Remarkably, over 70\% of the keypoints attain a PCK@50 score surpassing 96\%.
For the average metrics, PCK@20 and PCK@50 reach 63.47\% and 96.27\%, respectively.
Notably, the PowerSkel employs a CSI sensor with a sampling rate of only 30 Hz, considerably lower than the sensors utilized in prior literature~\cite{wang2019joint, wang2021point, zhou2023metafi++} in terms of both frequency and CSI dimensions. 
The Kinect's placement on one side to enhance visibility during power operations results in generally better performance in estimating keypoints on the left side of the human body compared to the right side. 
In some cases, right-side keypoints are obscured by the body, resulting in their absence. 
The uneven distribution of samples between missing and non-missing keypoints contributes to the lower PCK scores on the right side. 
Nevertheless, the CKDformer still exhibits commendable performance in estimating keypoints on the right side, and further improvements could be achieved with more comprehensive keypoints data.
Fig. \ref{fig_7} presents visualizations of test data for three typical actions: reversing the flow of power operations, swinging the handcart, opening and closing the cabinet door, as well as operating the knife gate.
In Fig. \ref{fig_7}, we compare the results of the CKDformer with Kinect V2.0 for human pose estimation. 
Fig. \ref{fig_7}(a) represents the ground truth obtained with Kinect V2.0. 
Fig. \ref{fig_7}(b) showcases the pose estimation results generated by CKDformer. 
CKDformer aligns closely with the ground truth, effectively capturing power operations via CSI-based pose estimation. 
The CKDformer's performance is on par with vision-based methods. 
We collected power operation data to evaluate our approach's effectiveness for power scenarios in low-light environments.
The embedded IR Emitter of Kinect V2.0 precisely captures the human pose without relying on external lighting, making it an ideal choice for annotating power operations in dark settings.

In Fig. \ref{fig_8}, we compare the results of different pose estimation approaches. 
Fig. \ref{fig_8}(a) shows the pose recognition results of Kinect V2.0. 
In contrast, Fig. \ref{fig_8}(b) represents an estimation based on OpenPose, which provides a rough estimate of the human form but needs more accuracy. 
Fig. \ref{fig_8}(c) presents results obtained by applying the online face++ skeleton detection model of MEGVII to input images, showcasing the limitations of vision-based methods in low-light conditions. 
In contrast, Fig. \ref{fig_8}(d) illustrates the visualization of PowerSkel, which remains unaffected by lighting conditions and consistently delivers accurate human pose estimations.
However, Kinect-based approaches face limitations in line-of-sight environments and high costs. The CKDformer excels in high-precision human pose estimation for power operations, addressing critical safety concerns in power scenarios.
The cost and functionalities of commonly utilized monitoring equipment are detailed in Table II. The prevailing monitoring equipment in power stations predominantly comprises natural light cameras, boasting a significant price advantage over other camera devices—an essential consideration for the power department. 
\begin{table}[!t]
\centering
\caption{Performance comparison on our dataset.}
\label{tab3}
\scalebox{0.9}{
\begin{tabular}{cc|ccccc}
\hline
\multicolumn{2}{c|}{Module}       & \multicolumn{5}{c}{Performance (\%)}                                                                                                \\ \hline
\multicolumn{1}{c}{SAF} & CKD    & \multicolumn{1}{c}{PCK@10} & \multicolumn{1}{c}{PCK@20} & \multicolumn{1}{c}{PCK@30} & \multicolumn{1}{c}{PCK@40} & PCK@50 \\ \hline
                    &   & 51.50                       & 72.06                       & 82.71                       & 88.77                       & 92.01  \\
$\surd$                        &        & 55.76                       & 76.97                       & 86.98                       & 91.54                       & 93.92  \\
                         & $\surd$      & 57.14                       & 78.19                       & 87.08                       & 91.51                       & 93.85  \\
$\surd$                        & $\surd$      & \textbf{63.47}                       & \textbf{83.49}                       & \textbf{91.48}                       & \textbf{94.74}                       & \textbf{96.27}  \\ \hline
\end{tabular}
}
\end{table}
However, natural light cameras face limitations in handling dark or poorly lit conditions and are constrained by the blind sector. Similarly, CSI-based devices such as Intel 5300NIC and Atheros come at a higher cost due to their reliance on host computers. In contrast, the ESP 32 SoC-based device proposed in this paper exhibits a substantial advantage over other devices in cost and functionality.

\subsection{Ablation Study}
Ablation experiments are conducted to assess the impact of the proposed SAF and CKD mechanisms, with the performance of each module detailed in TABLE III. 
The baseline, trained using raw data with a single Conformer, is a reference. 
Post-processing raw data with SAF demonstrated improved skeleton estimation, evidenced by a 4.26\% increase in PCK@10. 
Integrating SAF processing in models employing the CKD mechanism elevated PCK@10 from 57.14\% to 63.47\%. 
These results underscore SAF's role in denoising CSI data and refining features in power scenarios where CSI is susceptible to noise.

Analysis of CKD ablation results revealed the mechanism's efficacy in enhancing PCK performance. Without SAF, PCK@10 with CKD improved from 51.50\% to 57.14\%. 
\begin{figure}[!t]
\centering
\includegraphics[width=3.5in]{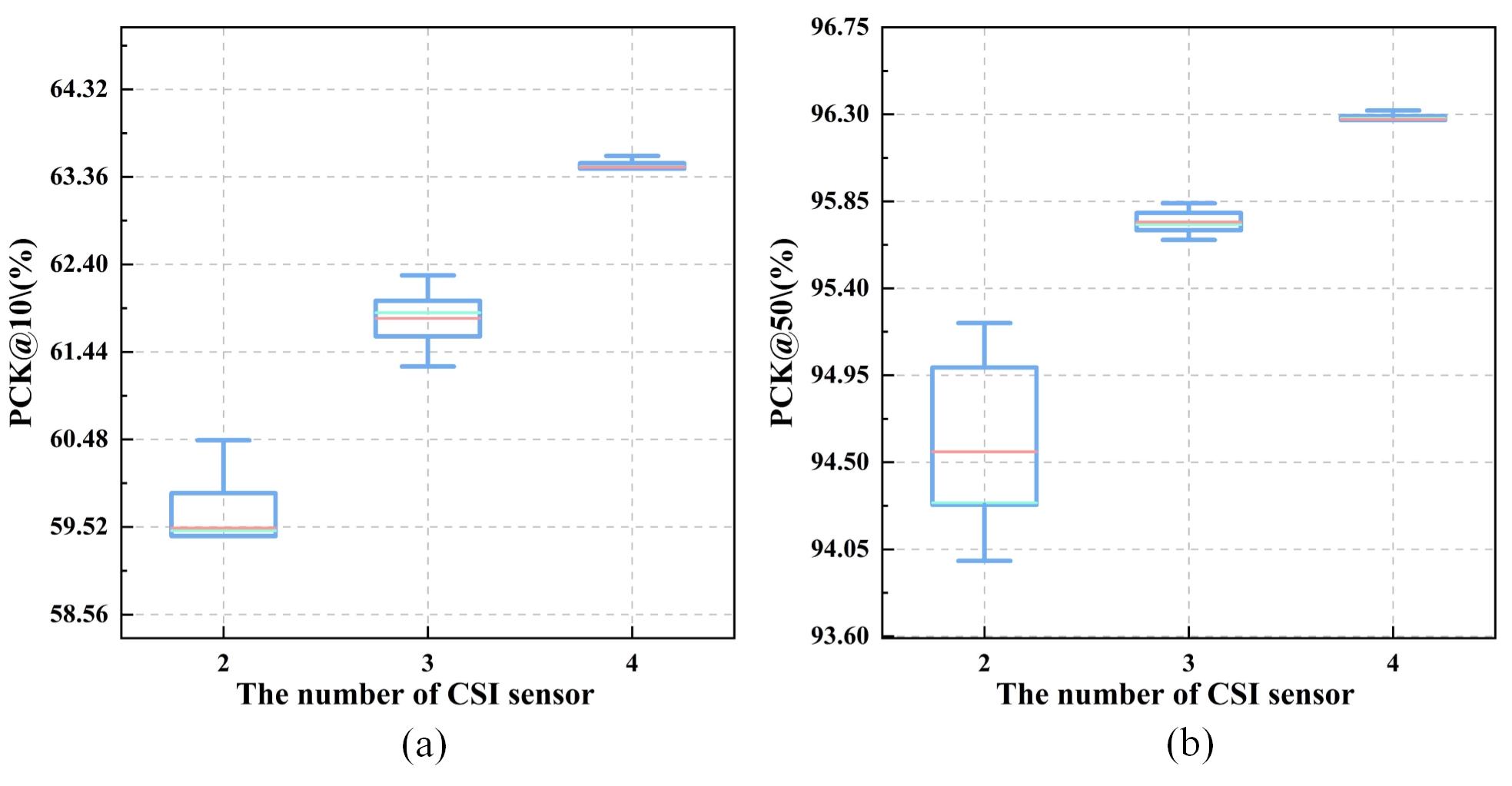}
\caption{Comparison of different sensor numbers. (a) Performance at PCK@10. (b) Performance at PCK@50.}
\label{fig_10}
\end{figure}
\begin{table}[!t]
\centering
\caption{Results of models with different architectures.}
\label{tab4}
\scalebox{0.85}{
\begin{tabular}{c|cccccc}
\hline
Model         & PCK@10 & PCK@20 & PCK@30 & PCK@40 & PCK@50 & \begin{tabular}[c]{@{}c@{}}model\\ size\end{tabular}  \\ \hline
Conf (1)       & 55.76  & 76.97  & 86.98  & 91.54  & 93.92  & 33.23       \\
ns-CKD        & 63.66  & 83.76  & 91.64  & 94.82  & 96.33  & 66.68       \\
CKD (2s)       & 63.47  & 83.49  & 91.48  & 94.74  & 96.27  & 33.35       \\
CKD (3s)       & 63.45  & 83.53  & 91.39  & 94.56  & 96.12  & 33.76       \\
CKD (4s)       & 63.34  & 83.29  & 91.22  & 94.52  & 96.04  & 33.82       \\ \hline
\end{tabular}
}
\end{table}
Remarkably, CKD improved PCK@10 by over 7\%, indicating its capacity to complement models and enhance estimation accuracy. 
Moreover, CKD exhibited enhanced performance when combined with SAF augmentation. 
The ablation study demonstrates the significant contributions of both SAF and CKD components to model performance, ultimately benefiting the accuracy of power operation estimation.

\subsection{Analysis on the Number of Sensors}
Experiments are conducted to assess the influence of CSI sensor number and position on pose estimation performance. 
In the context of the distribution room, spatial limitations and communication constraints led to a restriction on the number of sensors, allowing for a maximum of four sensors and a minimum of two sensors to ensure pairwise sensing. 
The experiments encompassed configurations and groupings of 2 and 3 sensors, as illustrated in Fig. \ref{fig_5}.
As shown in Fig. \ref{fig_10}, box plots showcasing the outcomes of PCK@10 and PCK@50, reveal the enhancement in pose estimation accuracy with increased sensors. 
Notably, using 2 sensors resulted in a 3.96\% reduction in PCK@10 compared to when 4 were employed. 
Various sensor combinations produced more pronounced differences in results, underscoring the significant impact of sensor spatial distribution on local feature capture. 
The CKDformer, built on the Conformer-based architecture, leverages these spatial characteristics.
The study underscores the substantial influence of both sensor number and position on pose estimation accuracy. 
However, it is essential to consider practical factors such as space constraints, cost, and communication limitations in power stations. 
Determining the optimal sensor position remains a focal point for future work.

\subsection{Cost Analysis with CKDformer}
We introduce a novel CSI acquisition method, PowerSkel, designed to enhance the richness of CSI data in power stations.
\begin{figure}[t]
\centering
\includegraphics[width=2.45 in]{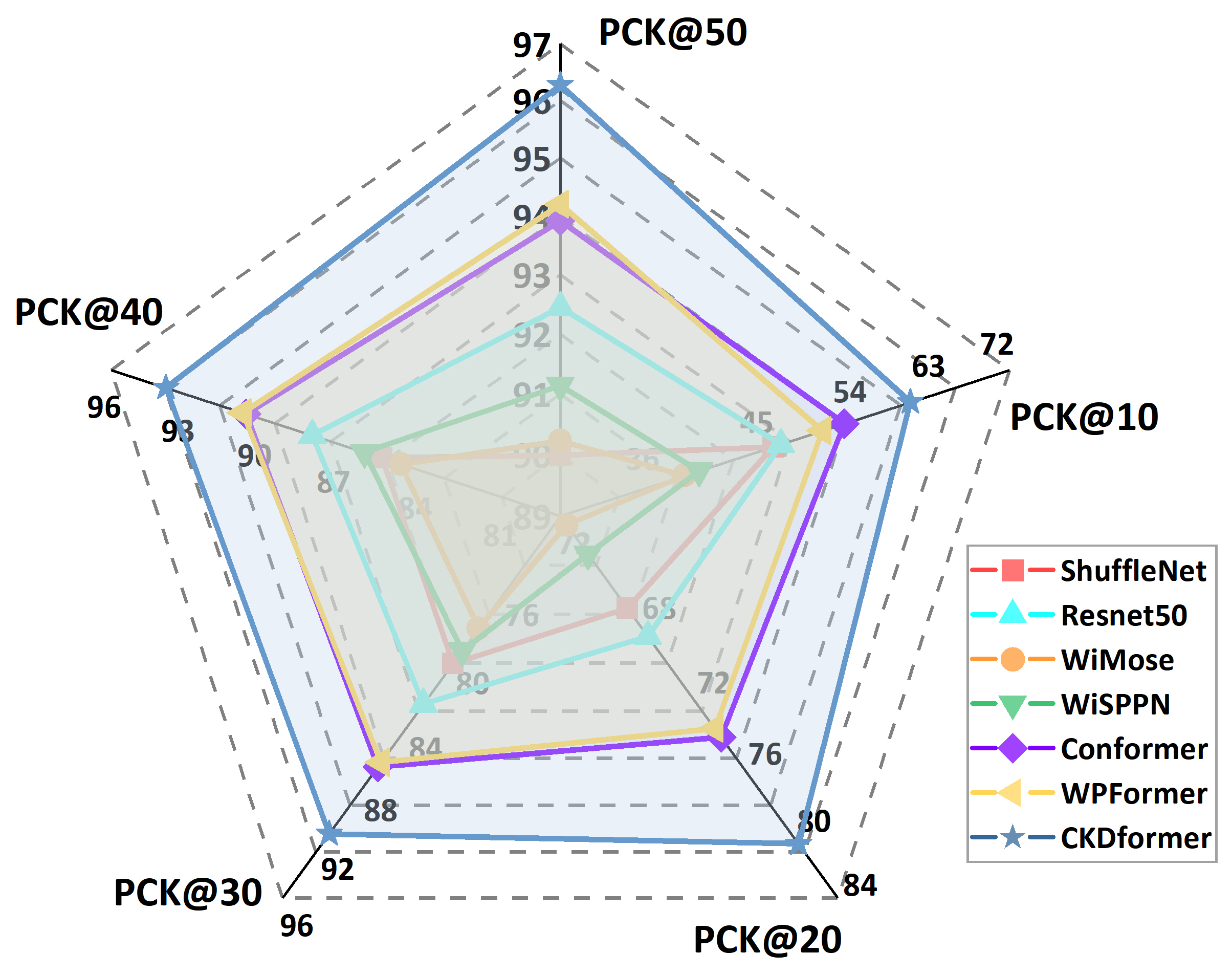}
\caption{Comparison of radargrams for different methods.}
\label{fig_11}
\end{figure}
\begin{table}[!t]
\centering
\caption{Performance comparison on our dataset.}
\label{tab5}
\scalebox{0.85}{
\begin{tabular}{c|ccccc}
\hline
Method     & PCK@10 & PCK@20 & PCK@30 & PCK@40 & PCK@50 \\ \hline
ShuffleNet\cite{zhang2018shufflenet} & 48.52  & 69.75  & 80.46  & 86.37  & 89.99  \\
Resnet50\cite{theckedath2020detecting}   & 49.24  & 71.26  & 82.97  & 88.98  & 92.45  \\
WiMose\cite{wang2021point}     & 40.79  & 65.44  & 78.37  & 85.66  & 90.23  \\
WiSPPN\cite{wang2019joint}     & 40.79  & 65.44  & 78.37  & 85.66  & 90.23  \\
Conformer\cite{gulati2020conformer}  & 55.76  & 76.97  & 86.98  & 91.54  & 93.92  \\
WPFormer\cite{zhou2023metafi++}   & 53.42  & 76.45  & 86.70  & 91.67  & 94.20  \\
\textbf{CKDformer}  & \textbf{63.47}  & \textbf{83.49}  & \textbf{91.48}  & \textbf{94.74}  & \textbf{96.27}  \\ \hline
\end{tabular}
}
\end{table}
The core innovation of CKDformer lies in maximizing channel relationships to enhance spatial CSI features. 
The effectiveness of CKDformer is substantiated through a series of comparative experiments presented in Table IV. 
These experiments encompass the evaluation of different models, including a single Conformer model (Conf (1)), a non-shared backbone CKDformer (nsCKD), CKDformer with 2 (CKD (2s)), 3 (CKD (3s)), and 4 (CKD (4s)) student models. 
The results indicate that CKDformer significantly outperforms the single Conformer model, demonstrating an approximate 2.5\% improvement in PCK@50. 
Part C of the results highlights the distinctive distribution of CSI spatial features, emphasizing the limitations of relying on single models for feature extraction. 
CKDformer addresses this issue by employing multiple student models to extract spatial CSI features from diverse perspectives, resulting in finer-grained feature representation through knowledge sharing. 
It's worth noting that the size of the non-shared ns-CKD model is twice that of CKD, reaching 66.68 M, while the size of CKD is 33.35 M. 
This discrepancy underscores the cost-efficient advantage of the shared backbone structure integrated into CKDformer. 
However, it's observed that beyond CKD (2s), the performance improvement in PCK doesn't continue with CKD (3s) and CKD (4s), suggesting potential overfitting due to the increased number of student models. 
Additionally, the exponential growth in model size for ns-CKD, which lacks a shared backbone, underscores the computational demands of accommodating more student models. 
These experiments reaffirm the practical significance of mining CSI features in power scenarios using CKDformer, which leverages inter-CSI channel data for effective knowledge sharing. 
The integration of a shared backbone structure not only enhances human pose estimation performance for power operations but also substantially reduces computational costs.

\subsection{Performance Comparisons among Different Approaches}
We evaluate the performance of CKDformer in power operator pose estimation by comparing it with six state-of-the-art models: ShuffleNet~\cite{zhang2018shufflenet} and ResNet 50~\cite{theckedath2020detecting} (both as CKDformer backbones), Conformer~\cite{gulati2020conformer}, WiMose~\cite{wang2021point}, WiSPPN~\cite{wang2019joint}, and WPFormer~\cite{zhou2023metafi++}. 
Although the concept of CSI-based pose estimation for power operations is introduced for the first time, we still conduct a comparative analysis with existing CSI-based pose estimation methods, i.e., WiMose, WiSPPN, and WPFormer. 
These methods are initially designed for monitoring human activities in home or office environments. 
Leveraging the robust image feature extraction capabilities of ShuffleNet and Resnet 50, known for their strong performance, we employ them as a backbone for CKDformer, replacing the Conformer for comparative evaluation. 
It is noteworthy that throughout this comparison, the inputs and outputs of all models, including CKDformer, remain consistent, comprising CSI, and skeletal coordinates obtained from Kinect.
All models are tested on the same dataset. Fig. \ref{fig_11} and Table V show that CKDformer outperforms the baselines significantly for CSI-based power operation pose estimation. 
CKDformer achieves a 2.07\% improvement in PCK@50 over the previous best model, WPFormer, demonstrating the effectiveness of its collaborative KD and self-attention structure. 
Moreover, CKDformer, ShuffleNet, and ResNet 50 have better results in PCK@10 than WiMose and WiSPPN, indicating that the cooperative learning mechanisms of CKDformer enhance accuracy.
Traditional single models struggle to handle complex power operation scenarios and fully exploit fine-grained CSI features. 
Due to their self-attention mechanisms, CKDformer, Conformer, and WPFormer perform better than other models. 
CKDformer surpasses convolution-based ResNet (limited to local features) and self-attention-based WPFormer (constrained by single-model limits). 
Knowledge sharing and the Conformer architecture of CKDformer balance global and local CSI features effectively, overcome single-model limitations, and mine inter-CSI channel features deeply. 
The comparative analysis confirms the superior accuracy of CKDformer in power operation pose estimation over other models. 
CKDformer establishes itself as a precise state-of-the-art solution for CSI-based pose estimation in power operation.

\subsection{Limitations and Future Work of PowerSkel}
As mentioned in the results of Part B, the limitation of PowerSkel lies in the accuracy of human skeletal labels during data acquisition. 
The visualization of PowerSkel can be improved by capturing precise and comprehensive keypoint labels. 
Power operation estimation presents unique challenges distinct from human pose estimation, notably the obstacle of potential obscuration caused by cabinet doors and other barriers. Even though we have demonstrated the effectiveness of CSI in detecting human pose behind obstacles with some experiments. More experiments are still needed to prove its feasibility in power operation. The environment change causes the distribution of CSI shifts, and the original Kinect labels become invalid. The environment shift is a typical cross-domain issue of CSI. Cross-domain is an important research direction in power operation based on CSI, and we will address this problem in future work.

\section{CONCLUSION}
In this paper, a CSI-based pose estimation approach for power operations, PowerSkel, is proposed for application in power station environments. 
PowerSkel leverages self-developed CSI sensors to create a sensing network and a CSI acquisition scheme that is specific to power scenarios, making it more cost-effective and easier to deploy. 
The CKDformer in PowerSkel extracts features from CSI and establishes the mapping relationship between CSI and keypoints to enable accurate pose estimation. 
Experimental results show that our method achieves 96.27\% of PCK@50, making it suitable even in dark environments. 
PowerSkel is a low-cost and high-precision solution that offers a novel way to estimate poses in power operations.

 
%

\newpage

\begin{IEEEbiography}[{\includegraphics[width=0.9in,clip,keepaspectratio]{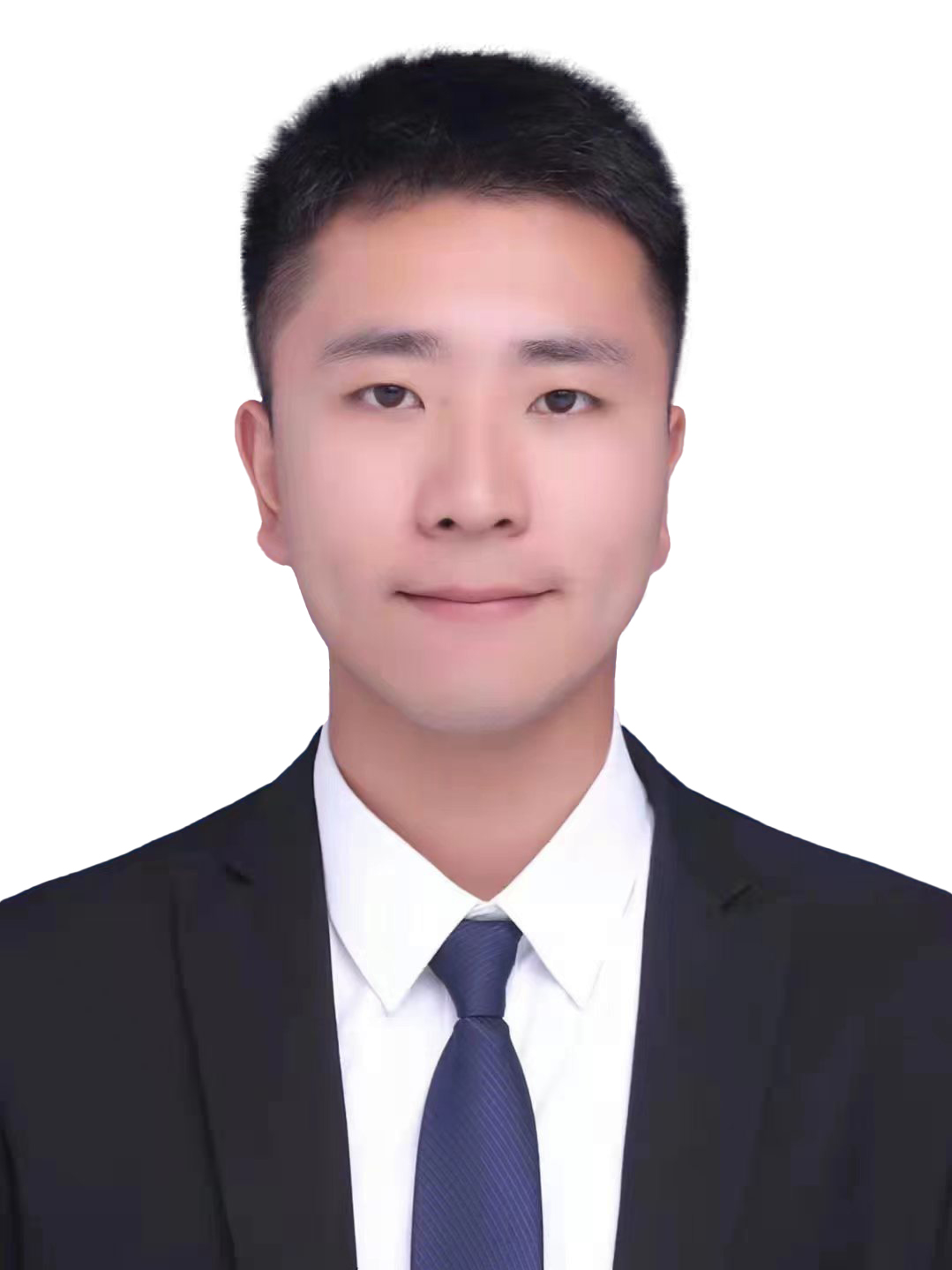}}]{Cunyi Yin}
received the B.S. degree from School of Control Engineering, Chengdu University of Information Technology, Chengdu, China, in 2016, and the M.S. degrees from Fuzhou University, Fuzhou, China, in 2020. He is pursuing the Ph.D. degree in  electric machines and electric apparatus with Fuzhou University, Fuzhou, China. Currently, he is engaged in joint research and cultivation in Machine Intellection Department, Institute for Infocomm Research, Agency for Science, Technology and Research (A*STAR), Singapore. His current research interests include electrical safety, human activity recognition, internet of things, wireless sensing and machine learning.
\end{IEEEbiography}

\begin{IEEEbiography}[{\includegraphics[width=0.9in,clip,keepaspectratio]{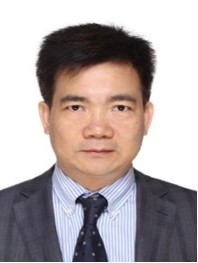}}]{Xiren Miao}
received the B.S. degree from Beihang University, Beijing, China, in 1986, and the M.S. and Ph.D. degrees from Fuzhou University, Fuzhou, China, in 1989 and 2000, respectively, where he is currently a Professor with the College of Electrical Engineering and Automation. His research interests include human activity recognition, electrical and its system intelligent technology, and diagnosis of electrical equipment.
\end{IEEEbiography}

\begin{IEEEbiography}[{\includegraphics[width=0.9in,clip,keepaspectratio]{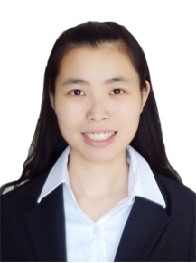}}]{Jing Chen}
received the B.S., M.S., and Ph.D. degrees from Xiamen University, Fujian, China, in 2010, 2013, and 2016, respectively. She is currently an Associate Professor with the College of Electrical Engineering and Automation, Fuzhou University. Her research interests include human activity recognition, intelligent sensor network and machine learning.
\end{IEEEbiography}

\begin{IEEEbiography}[{\includegraphics[width=0.9in,clip,keepaspectratio]{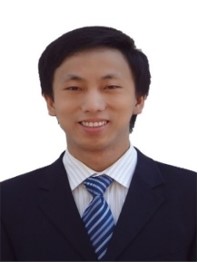}}]{Hao Jiang} received the B.S. degree in automation from Xiamen University, Fujian, China, in 2008, and the Ph.D. degree in control theory and control engineering from Xiamen University, Fujian, China, in 2013. He is currently an Associate Professor with the College of Electrical Engineering and Automation, Fuzhou University. His research interests include the internet of things, human activity recognition, artificial intelligence and machine learning.
\end{IEEEbiography}

\begin{IEEEbiography}[{\includegraphics[width=0.9in,clip,keepaspectratio]{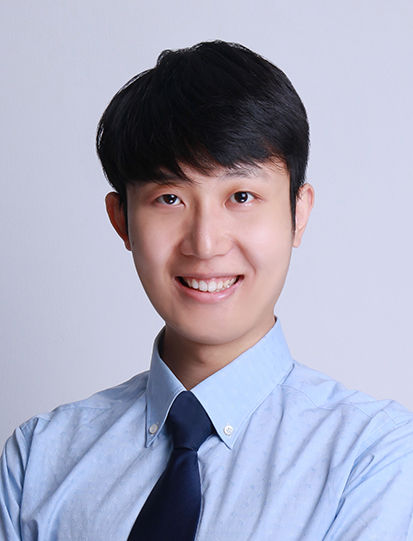}}]{Jianfei Yang} received the B.Eng. from the School of Data and Computer Science, Sun Yat-sen University in 2016, and the Ph.D. degree from Nanyang Technological University (NTU), Singapore in 2021. He used to work as a senior research engineer at BEARS, the University of California, Berkeley. His research interests include deep learning, wireless sensing, and artificial intelligence of things. He received the best Ph.D. thesis award from NTU, and won many International AI challenges in computer vision and interdisciplinary research fields. Currently, he is a Presidential Postdoctoral Research Fellow and an independent PI at NTU.
\end{IEEEbiography}

\begin{IEEEbiography}[{\includegraphics[width=0.9in,clip,keepaspectratio]{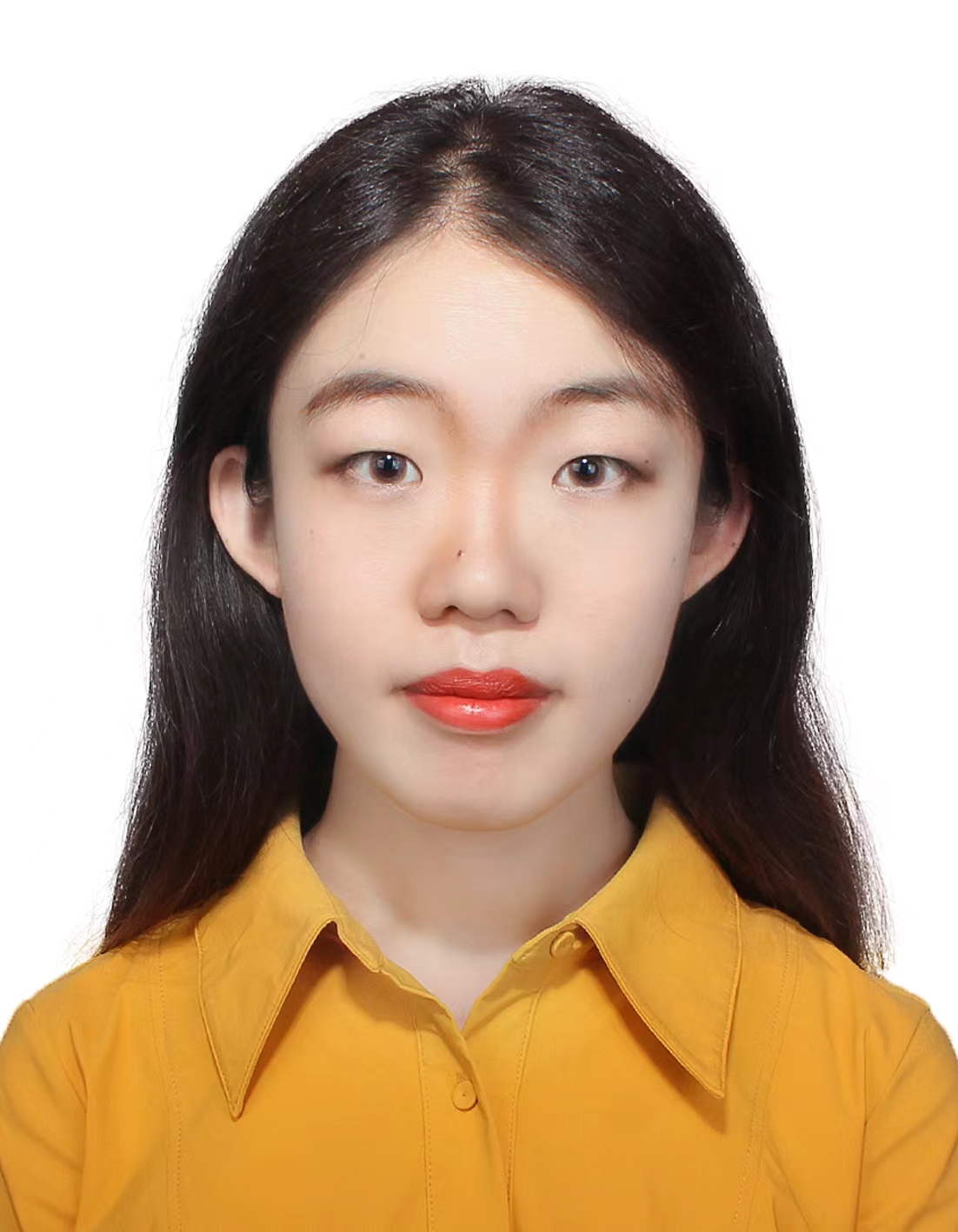}}]{Yunjiao Zhou} received the B.Eng. from the School of Information Science and Technology, Southwest Jiaotong in 2021, and the Msc. degree in Electrical and Electronic Engineering from Nanyang Technological University in 2022. She is currently pursuing a Ph.D. degree with the school of Electrical and Electronic Engineering in Nanyang Technological University. Her research interests include multi-modal learning for multi-sensor applications and human perception in smart home.
\end{IEEEbiography}

\begin{IEEEbiography}[{\includegraphics[width=0.9in,clip,keepaspectratio]{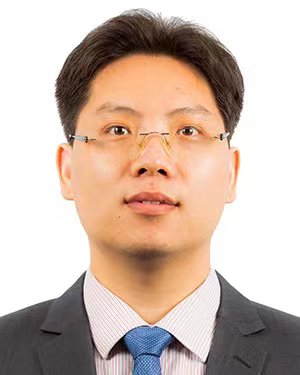}}]{Min Wu} received the B.S. degree in computer science from the University of Science and Technology of China (USTC), Hefei, China, in 2006, and the Ph.D. degree in computer science from Nanyang Technological University (NTU), Singapore, in 2011. He is currently a senior scientist in Machine Intellection Department, Institute for Infocomm Research, Agency for Science, Technology and Research (A*STAR), Singapore. His current research interests include machine learning, data mining and bioinformatics. He received the best paper awards in InCoB 2016 and DASFAA 2015, and the finalist academic paper award in IEEE PHM 2020. He also won the CVPR UG2+ challenge in 2021 and the IJCAI competition on repeated buyers prediction in 2015.
\end{IEEEbiography}

\begin{IEEEbiography}[{\includegraphics[width=0.9in,clip,keepaspectratio]{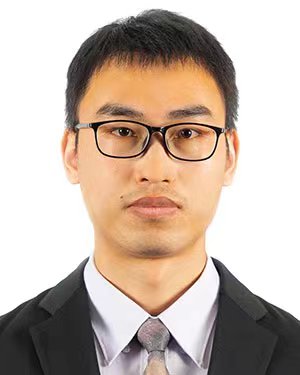}}]{Zhenghua Chen} received the B.Eng. degree in mechatronics engineering from University of Electronic Science and Technology of China (UESTC), Chengdu, China, in 2011, and Ph.D. degree in electrical and electronic engineering from Nanyang Technological University (NTU), Singapore, in 2017. Currently, he is a Scientist, PI and Lab Head at Institute for Infocomm Research, and an Early Career Investigator at Centre for Frontier AI Research (CFAR), Agency for Science, Technology and Research (A*STAR), Singapore. His research interests include data-efficient and model-efficient learning with related applications in smart city and smart manufacturing. He has won several competitive awards, such as First Place Winner for CVPR 2021 UG2+ Challenge, A*STAR Career Development Award, Norbert Wiener Review Award, Best Paper Award at IEEE ICIEA 2022 and IEEE SmartCity 2022, etc. He serves as Associate Editor for IEEE Transactions on Industrial Informatics, IEEE Transactions on Instrumentation and Measurement, and Elsevier Neurocomputing. He is currently the Chair of IEEE Sensors Council Singapore Chapter and IEEE Senior Member.
\end{IEEEbiography}


 




\vfill

\end{document}